\documentclass[aps,prl,twocolumn,amsmath,amssymb,superscriptaddress]{revtex4-2}

\usepackage{graphicx}
\usepackage{color}
\definecolor{cerulean}{rgb}{0.0, 0.48, 0.65}
\definecolor{regalia}{rgb}{0.32, 0.18, 0.5}
\usepackage{float}
\usepackage{accents}
\usepackage{hyperref}
\usepackage{bbold}
\usepackage{soul}
\hypersetup{
    colorlinks=true,
    linkcolor=regalia,
    filecolor=regalia,      
    urlcolor=regalia,
    citecolor=regalia
}

\usepackage[resetlabels, labeled]{multibib}

\newcites{S}{Supplemental material}

\usepackage[normalem]{ulem}
\usepackage{subfigure}
\usepackage{tikz} 
 \usetikzlibrary{decorations.pathreplacing,calligraphy}

\def\be{\begin{equation}}
\def\ee{\end{equation}}
\def\ba{\begin{eqnarray}}
\def\ea{\end{eqnarray}}
\def\bc{\begin{center}}
\def\ec{\end{center}}

\newcommand{\gsim}{\raisebox{-0.13cm}{~\shortstack{$>$ \\[-0.07cm]
      $\sim$}}~}

\begin{document}
\title{Quantum cellular automata for quantum error correction and density classification}

\author{T. L. M. Guedes}
\affiliation{Institute for Quantum Information, RWTH Aachen University, D-52056 Aachen, Germany}
\affiliation{Peter Gr{\"u}nberg Institute, Theoretical Nanoelectronics, Forschungszentrum J{\"u}lich, D-52425 J{\"u}lich, Germany}

\email{t.guedes@fz-juelich.de}

\author{D. Winter}

\affiliation{Institute for Quantum Information, RWTH Aachen University, D-52056 Aachen, Germany}
\affiliation{Peter Gr{\"u}nberg Institute, Theoretical Nanoelectronics, Forschungszentrum J{\"u}lich, D-52425 J{\"u}lich, Germany}

\author{M. M{\"u}ller}

\affiliation{Institute for Quantum Information, RWTH Aachen University, D-52056 Aachen, Germany}
\affiliation{Peter Gr{\"u}nberg Institute, Theoretical Nanoelectronics, Forschungszentrum J{\"u}lich, D-52425 J{\"u}lich, Germany}

\begin{abstract}

Quantum cellular automata are alternative quantum-computing paradigms to quantum Turing machines and quantum circuits. Their working mechanisms are inherently automated, therefore measurement free, and they act in a translation invariant manner on all cells/qudits of a register, generating a global rule that updates cell states locally, i.e., based solely on the states of their neighbors. Although desirable features in many applications, it is generally not clear to which extent these fully automated discrete-time local updates can generate and sustain long-range order in the (noisy) systems they act upon. In special, whether and how quantum cellular automata can perform quantum error correction remain open questions. We close this conceptual gap by proposing quantum cellular automata with quantum-error-correction capabilities. We design and investigate two (quasi-)one dimensional quantum cellular automata based on known classical cellular-automata rules with density-classification capabilities, namely the local majority voting and the two-line voting. We investigate the performances of those quantum cellular automata as quantum-memory components by simulating the number of update steps required for the logical information they act upon to be afflicted by a logical bit flip. The proposed designs pave a way to further explore the potential of new types of quantum cellular automata with built-in quantum-error-correction capabilities. 

\end{abstract}

\maketitle

\emph{Introduction.}---Cellular automata (CAs) were proposed as simplified models of self-reproducing systems~\cite{NeumannCA, game}, but rapidly grew into powerful paradigms for the description of complex systems constructed from identical components with simple and local interactions~\cite{wolfram_complex}. 
Such emerging complexity renders CAs suitable candidates for computers~\cite{wolfram_comp}, with proven universality~\cite{Smith1, Smith2} and reversibility (i.e., any irreversible CA can be simulated by a reversible CA)~\cite{ToffoliCA, Morita}. Error correction (EC) with CAs has been studied as a density-classification problem, i.e., whether a CA can force all cells of the system to the state that the majority of cells in any given initial configuration were in. It has been shown that no CA with two states per cell can perfectly classify the density when the number of cells is sufficiently large~\cite{nodcp1, nodcp2}. Nonetheless, a specific combination of CAs generates a perfect density classifier (DC) in the absence of noise~\cite{Fuks, Mendonca}, and certain CAs display formidable performances as DCs, even in the presence of noise~\cite{GKL, Toom, Park, Gacsproof1, Gacsproof2, Gray}.

Quantum cellular automata (QCAs), the quantum counterparts of CAs~\cite{Feynman}, were proposed as alternative universal quantum-computation models to quantum circuits and quantum Turing machines~\cite{Watrous, program_QCA}. QCAs are defined axiomatically~\cite{Schumacher, Richter} and can be cast as local finite-depth circuits~\cite{Arrighi_unitarity, Arrighi}. In QCAs, each cell is a quantum subsystem and the total system evolves unitarily in discrete time steps in a translation-invariant and quantum-locality-preserving (also called causal) manner~\cite{ArrighiQED1, MlodinowQED, otherQED}. This means that local operators are mapped into quasi-local operators at each step~\cite{Farrelly, Arrighi, Perez, time_asymptotics}. By implementing the evolution as an automorphism, QCAs can bypass the requirement of single/few-qubit addressability, a prominent feature in proposals for experimental implementations of QCAs~\cite{experiment}. Even with QCAs making their way towards experimental realizations~\cite{DCAexp, newexperiment}, it remains unclear if and how the quantum equivalent of the density-classification problem, quantum error correction (QEC), can be integrated in this framework, even though (classical) CAs have already been proposed to automatize syndrome analysis in QEC~\cite{Buechler, Harrington, Terhal, Fibonacci, fieldCA2,fieldca, Kubica1, Kubica2, Kubica3}.

\begin{figure*}
    \centering
    \includegraphics[width=\linewidth]{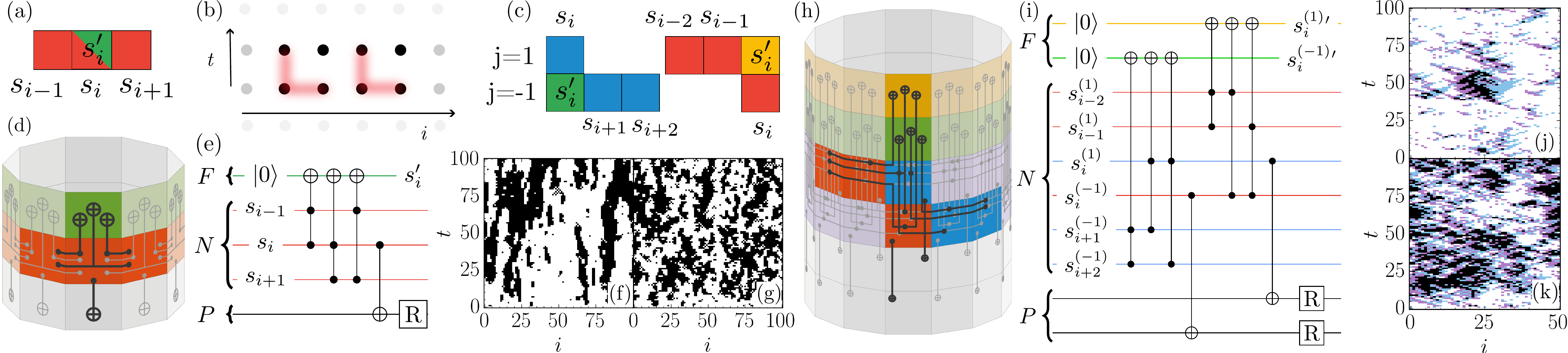}
 \caption{Representations of Q232 (left) and QTLV (right). To guarantee reversibility, the QCAs are extended in one additional timelike dimension [vertical axes in the cylinders (d) and (h)], so that the global rule, shown as a combination of commuting and translation-invariant local unitaries $U_i$ (colored), acts on the present register, covering an entire time-$t$ section of the cylinder, while coupling to the future register (above) and decoupling from the past register (below). A single $U_i$ (one for each $j$ in the case of QTLV) is highlighted in each cylinder, showing how these local unitaries are decomposed into geometrically distributed parallelizable quantum gates, depicted in standard form in (e) for Q232 and (i) for QTLV, where $P,N,F$ denote respectively past, present/now and future, while $\mathrm{R}$ stands for reset. The classical neighborhood schemes $N_d$ of the corresponding CAs are also highlighted in red [for Q232 in (d)] and blue and red [for QTLV in (h)], and they are also shown in isolated form in (a) and (c); in these representations, the future cells are shown in green (Q232) or green and yellow (QTLV) and their states are marked by primes. (f) and (j) show 232 and TLV orbits, respectively, under noise with bit-flip probability $p=1/12$, while (g) and (k) show orbits when $p=1/6$. In (j) and (k), each bistring is shown as a single string with 4 states: 00 (white), 01 (blue), 10 (purple) and 11 (black). Lastly, (b) represents schematically a potential Q232 realization in arrays of Rydberg atoms.}
  \label{fig1}
\end{figure*}

In this Letter, we study the density-classification performance under noise of two (quasi-)1D CA rules, Wolfram's rule 232~\cite{Wolfram_statistical} (also known as local majority voting) and Toom's two-line voting (TLV)~\cite{Toom}. We then propose QCAs corresponding to these rules and translate them into quantum circuits (cf. Fig~\ref{fig1}). Those are not only the first quantum-error-correcting (or quantum-DC) QCAs proposed, but, when concatenated, also proofs of the feasibility of QEC in or with QCA architectures. Their working principle is fundamentally different from usual QEC, since no measurements are needed~\cite{measurefree1, measurefree2, measurefree3, measurefree4, measurefree5, measurefree6, David, Igor, Saffman_free, Sascha}. Consequently, syndrome collection and classical decoding are absent, making our QCAs a fully quantum approach to QEC without quantum-to-classical interfaces. 
We simulate our QCAs' performances in the presence of coherent and incoherent phenomenological bit-flip noise, as well as incoherent depolarizing circuit noise. Our simulations show that quantum two-line-voting (QTLV) is an excellent candidate for a quantum-memory component~\cite{memory}. We compare the performances of our (Q)CAs to global voting, i.e., the classical repetition code~\cite{Buechler}, and show that under certain circumstances QTLV outperforms the repetition code~\cite{repetition, google}. We also briefly discuss some possibilities for experimental implementations of our QCAs. Lastly, we discuss how full QEC can be enabled through concatenation of our designs.

\emph{(Classical) Cellular automata.}---A deterministic $d$-dimensional CA, $(L_d, S, f,N_d)$, is a system defined by a $d$-dimensional (often infinite) lattice $L_d$ of identical cells, each having an internal set of states $S$ and evolving according to a local function (rule) $f$ mapping a neighborhood $N_d$ of each cell into the update value of that cell~\cite{Morita}.  
A configuration over $S$ (global state) is a map $\zeta: L_d \to S$, such that $\zeta (i)=s_i$ for the cell state $s_{i}$ on site $i$. The set of all configurations is denoted $\text{Conf}(S, L_d)$ and the global transition rule is a function $F:\text{Conf}(S, L_d)\to \text{Conf}(S, L_d)$. Space-time configurations generated by $t$ applications of $F$ on an initial configuration $\zeta$, $F^{(t)}(\zeta)\equiv F\circ F \circ \ldots \circ F(\zeta)$, are called orbits [cf. Figs.~\ref{fig1}(f),(g),(j),(k)]. 

CA rules can be reversible or irreversible: every irreversible $d$-dimensional CA can be simulated by a reversible $(d+1)$-dimensional CA~\cite{ToffoliCA} and irreversible 1D CAs can always be simulated by some reversible 1D CA~\cite{Morita}. 

Although CAs cannot be perfect DCs~\cite{nodcp1, nodcp2}, Ref.~\cite{Fuks} showed that a combination of rules 184 and 232 in $(L_1=\mathbb{Z}_n, S=\mathbb{Z}_2)$ systems 
generates a perfect DC. Rule 184, however, does not fulfill self-duality~\cite{Buechler}, 
what hinders the decoupling between different-time configurations in the corresponding QCA, as discussed later. We therefore initially focus on  rule 232. We assume periodic boundary conditions throughout, since they best emulate the infinite-lattice scenario often encountered in CA/QCA theory and lead to superior performances in DCs. Other boundary conditions are also possible~\cite{Buechler}.

Rule 232 can be described as
\begin{equation}
    s_i \to s_{i+1} s_{i-1} + s_{i+1}s_i + s_i s_{i-1} \mod 2
    \label{maj}
\end{equation}
and imprints on the central cell the majority (modulo 2) of all states $s_i\in \{0,1\}$ in the  neighborhood [see Fig.~\ref{fig1}(a)]. As a noiseless DC, it can correct sole errors (e.g., a single state-1 cell in a background of state-0 cells or vice-versa), and as long as sole errors are sparsely distributed throughout the lattice, successful density classification is possible. However, when errors group in the form of islands, rule 232 will map the latter into themselves, therefore making density classification impossible. This explains the poor performance of rule 232 under noise: Whenever neighboring errors are created, density classification fails. In the presence of noise, 2- and 3-error islands have higher probabilities to grow than to shrink, while larger islands have similar probabilities for both cases. Therefore, over time the islands will contain over half of the cells~\cite{Suppl_Mat}. 

A far more powerful quasi-1D ($\tilde{1}$D) DC, TLV, can be built from the 232 CA by extending the lattice to a double string, $L_{\tilde{1}}=\mathbb{Z}_{n/2}\oplus \mathbb{Z}_{n/2}$, and applying local majority voting, Eq.~\eqref{maj}, with two different neighborhoods, each updating the cells on a different string~\cite{Toom} [cf. Fig.~\ref{fig1}(c)]. Denoting the upper/lower string by superscripts $j=\pm 1$, TLV can be described as [cf. Fig.~\ref{fig1}(c)]
\begin{equation}
    s^{(j)}_i \to s^{(j)}_{i-j} s^{(j)}_{i-2j} + s^{(j)}_{i-j}s^{(-j)}_i + s^{(-j)}_i s^{(j)}_{i-2j} \mod 2 .
    \label{TLV}
\end{equation} 
In the absence of noise, it can be shown that every finite island of $2l+1$ errors on an infinite background of zeros or ones is eroded after a time $\leq ml$ for some CA-specific constant $m$, making TLV a linear eroder~\cite{Park}. We show below that even with noise this property guarantees good performances. 

\emph{Quantum cellular automata.}---A deterministic QCA, $(\Gamma_d, \mathcal{H}_i, \tilde{\mathcal{A}}_i, u, \mathcal{R}_i)$, is a system defined by a $d$-dimensional lattice $\Gamma_d$ of identical cells, each of which has a Hilbert space $\mathcal{H}_i$ with a corresponding observable algebra $\tilde{\mathcal{A}}_i$ for $i\in \Gamma_d$. The QCA Hilbert space is $\mathcal{H}=\otimes_{i\in \Lambda} \mathcal{H}_i$ and its observable algebra is $\mathcal{A}=\otimes_{i\in \Lambda} \tilde{\mathcal{A}}_i$, where $\Lambda = \Gamma_d$ ($\Lambda \subset \Gamma_d$) for (in)finite $\Gamma_d$\footnotetext[7]{It should be noted that an infinite tensor product of
  Hilbert spaces is an ill-defined construct. For this reason, the QCA Hilbert
  space can only be defined as $\protect \mathcal {H}=\otimes _{i\in \Gamma _d}
  \protect \mathcal {H}_i$ if $\Gamma _d$ is finite. For an infinite lattice,
  the QCA Hilbert space has to be defined over sufficiently large regions
  $\Lambda $ to allow for a description close to the infinite case. This is
  exactly the reason why QCA evolution is described in terms of automorphisms
  rather than unitaries. It is also the reason why a QCA description in terms
  of an alphabet Hilbert space makes use of (infinitely many) quiescent states
  to allow for a finite number of non-quiescent cell states composing a finite
  number of allowed configurations~\cite {Arrighi}.}~\cite{Arrighi, Note7}. The QCA local operators are $a_i=\tilde{a}_i\otimes_{l\neq i}\tilde{\mathbb{1}}_l$ for $\tilde{a}_i\in \tilde{\mathcal{A}}_i$ and $l\in \Lambda$ and belong to the QCA local observable algebra $ \mathcal{A}_i\cong \tilde{\mathcal{A}}_i$. The QCA evolves in time steps according to an automorphism (global rule) $u:{\mathcal{A}}_i\to \mathcal{A}_{\mathcal{R}_i}$ mapping local observables $a_i$ of each cell $i$ into quasi-local observables $a_{\mathcal{R}_i}\in\mathcal{A}_{\mathcal{R}_i}\cong \otimes_{l\in\mathcal{R}_i} \tilde{\mathcal{A}}_l$ acting nontrivially within some region $\mathcal{R}_i\subset \Gamma_d$ surrounding $i$~\cite{infinite, Farrelly} (note that an invertible map like $u$ requires $\mathcal{A}_{\mathcal{R}_i}$ and $\mathcal{A}_{i}$ to cover the same set of cells, $\Lambda$). Besides being translation invariant, $u$ is locality preserving, since it prevents the observables from spreading through the entire lattice in a single time step~\footnote{In the language of QEC, this means that a physical error is
  not mapped into a logical error after a single time step.}. For finite lattices, $\exists \, U\in \mathcal{A}$ such that $u(a_i)=U^\dag a_i U$.  
Lastly, there is a one-to-one correspondence (wrapping lemma~\cite{Schumacher}) between any translation invariant QCA $(\Gamma_d, \mathcal{H}_i, \mathcal{A}_i, u, \mathcal{R}_i)$ on an infinite lattice and an equivalent QCA $(\Gamma'_d, \mathcal{H}_i, \mathcal{A}_i, u, \mathcal{R}_i)$ on a finite lattice with periodic boundary conditions. The two QCAs behave similarly provided that $\Gamma'_d$ is sufficiently larger than $\mathcal{R}_i$~\footnote{Namely, $\Gamma'_d$ should be large enough relative to $\mathcal{R}_i$ that the intersection between overlapping neighborhoods of two cells in $\Gamma_d$, when one of them gets translated by some periods of $\Gamma'_d$, becomes empty.}.

To achieve QCAs from the 232 and TLV CAs, we first make the latter reversible by encoding with $F$ the time-$(t+1)$ configuration in a new (future) register while the time-$t$ configuration remains stored in the present $N$ register~\cite{ToffoliCA}. By keeping information from the states in the $t$- and $(t+1)$-registers, one can construct reversible transition matrices that map $[s_{i_0,t},s_{i_1,t},s_{i_2,t},s_{i_3,t+1}] \to [s_{i_0,t},s_{i_1,t},s_{i_2,t},s_{i_3,t+1} + f(s_{i_0,t}, s_{i_1,t} , s_{i_2,t}) \mod 2]$ for time-$t$ input cells $i_0,i_1,i_2 \in L_d$ whose states give through rule $f$ the update ($t+1$) value of cell $i_3$~\cite{ToffoliCA, Suppl_Mat}. In our case, $f$ generates the transformations \eqref{maj} and \eqref{TLV} for 232 and TLV, respectively, and for TLV $i_0,i_1,i_2, i_3$ also contain information about the location in the upper or lower strings. Such constructions can be seen as extensions of $L_d$ to $L_{d+1}$ with the additional dimension representing time [cf. Figs.~\ref{fig1}(d) and (h)].

We consider each cell to be a qubit, and the QCAs to be started in an all-0 space-time configuration; the logical state is then encoded on the $t=0$ (bi)string. It turns out that self-duality in the CAs to be quantized, defined as $f(\neg s_{i-1}, \neg s_{i}, \neg s_{i+1})= \neg f(s_{i-1}, s_{i}, s_{i+1}) $, is a key property to allow for decoupling of past and present configurations. Self-duality means that two CA configurations $\zeta$ and $\bar{\zeta}$ with all cells in complementary states [i.e.,  $\zeta(i)=\bar{\zeta}(i)\oplus 1\mod 2\; \forall  i  \in L_d$] will conserve this symmetry throughout their orbits, so that flipping all past cells of the orbit of $\bar{\zeta}$ makes all past configurations of $\zeta$ and $\bar{\zeta}$ the same. For a QCA constructed from a self-dual CA, that means that the present configuration can be kept as a coherent logical superposition of $|\zeta^{(t)} \rangle =\hat{E}  \otimes_i |0_{i,t}\rangle$ and $ |\bar{\zeta}^{(t)} \rangle =\hat{E} \otimes_i |1_{i,t}\rangle $, with some set of errors $\hat{E}$ on both, while past configurations are decoupled as a product state (which is the same for both logical states), $|\bar{\zeta}^{(t')} \rangle \to |\zeta^{(t')} \rangle $ for $t'<t$. It is worth noting that naive quantization of CAs often violates unitarity, translation invariance, locality and/or self-duality; the range of CA-derived 1D QCAs that perform QEC is therefore constrained. Hamiltonian systems, having continuous-time dynamics and generating generally non-local unitaries, cannot be QCAs~\cite{Farrelly, NoQCA1, NoQCA2}.

The automorphism $u$ can be built from a product of (quasi-)local unitaries $U_{i}$ associated with cells $i\in \mathcal{R}'_i\subseteq \mathcal{R}_i$ such that $U^\dag_{i}(\cdot )U_i:  {\mathcal{A}}_{\mathcal{R}'_i} \to \mathcal{A}_{\mathcal{R}'_i}$~\cite{Schumacher}. In fact, to evolve the observables ${a}_i$ it suffices to apply a product of a few $U_j$, since for $ 2i -j\not\in \mathcal{R}'_i$ those unitaries act on the observable as the identity. $\mathcal{R}_i$ is then (contained in) the union of all $\mathcal{R}'_j$ such that $2i-j\in \mathcal{R}'_i$. For our quantized CAs, $\mathcal{R}'_i=N_{d+1}(i)$. In summary, $u({a}_i)= (\prod_j U_j)^\dag {a}_i (\prod_j U_j) |_{2i-j\in \mathcal{R}'_i}$ and, for finite $\Gamma_d$, $U=\prod_k U_k$. Such operator products are well defined only when $U_iU_k=e^{i\theta_{ik}}U_k U_i$ for some phase $\theta_{ik}$, and therefore commuting unitaries are part of our QCA designs.

\emph{Quantum local majority voting (Q232).}---The quantum version of rule 232, Q232, can be achieved by translating Eq.~\eqref{maj} into
\begin{equation}
   U_{i,t}=  (\sigma^{(x)}_{i,t-1})^{c_{i,t}}(\sigma^{(x)}_{i,t+1})^{[c_{i+1,t}c_{i-1,t}+c_{i+1,t}c_{i,t}+c_{i,t}c_{i-1,t}]}
    \label{qmaj}
\end{equation}
for $c_{i,t}=[\mathbb{1}_{i,t}-\sigma^{(z)}_{i,t}]/2$. Note that $\sigma^{(x)}_{i,t}= \exp [\pm i\pi (\mathbb{1}_{i,t}-\sigma^{(x)}_{i,t})/2]$, therefore~\eqref{qmaj} can be expressed as an exponential operator. The $t+1$ operator in Eq.~\eqref{qmaj} can be decomposed into a product of 3 commuting Hermitian unitaries, one for each term in its exponent, each of which corresponds to a Toffoli gate. Similarly, the $t-1$ operator is a CNOT gate. We therefore see that the application of Q232 corresponds locally to 3 Toffoli gates and one CNOT per cell in the lattice, as shown in Figs.~\ref{fig1}(d),(e). The Toffoli gates are responsible for encoding into the future cell $i$ the majority amongst the present-cell states in the neighborhood of $i$: $|s_{i-1,t},s_{i,t},s_{i+1,t}\rangle |0_{i,t+1}\rangle \to |s_{i-1,t},s_{i,t},s_{i+1,t}\rangle |s_{i,t+1}\rangle$ with $s_{i,t+1}$ given by the right-hand side of Eq.~\eqref{maj}. The CNOTs make use of self-duality to decouple the past cells from the present and future ones. In this way, Q232 keeps a coherent logical superposition of $2n$ qubits. To recover the $n$-qubit logical state/configuration, only the CNOTs are applied on the penultimate configuration. Similarly, given an initial logical state, the first application of Q232 uses only the Toffoli gates. One can also redefine Q232 so that the Toffoli gates are applied from $t\to t+1$ and the CNOTs afterwards from $t+1\to t$, so that a logical state of $n$ qubits is kept at each time step. 

\emph{Quantum two-line voting (QTLV).}---By including an additional string-related superscript $j=\pm 1$ in Eq.~\eqref{qmaj}, we derive the local-evolution unitary for QTLV:
\begin{equation}
   U_{i,t}^{(j)}=  (\sigma^{(x,j)}_{i,t-1})^{c^{(j)}_{i,t}}(\sigma^{(x,j)}_{i,t+1})^{[c^{(j)}_{i-j,t}c^{(j)}_{i-2j,t}+c^{(j)}_{i-j,t}c^{(-j)}_{i,t}+c^{(-j)}_{i,t}c^{(j)}_{i-2j,t}]}
    \label{qtlv}
\end{equation}
with $c^{(j)}_{i,t}=[\mathbb{1}^{(j)}_{i,t}-\sigma^{(z,j)}_{i,t}]/2$. For each $i$, $t$ and $j$, Eq.~\eqref{qtlv} can also be decomposed into 3 Toffoli gates and one CNOT [see Figs.~\ref{fig1}(h),(i)]. Since TLV works on two $n/2$-cell strings at each time step, QTLV acts on a total of 6 strings. One can, however, break each time step into two moves, one made of Toffoli gates ($t\to t+1$) and one made of CNOTs ($t+1\to t$), so that a total of 4 strings are used. The global rule and locality-preserving properties of Q232 and QTLV are discussed in~\cite{Suppl_Mat}.

\begin{figure}
    \centering
    \includegraphics[width=1\linewidth]{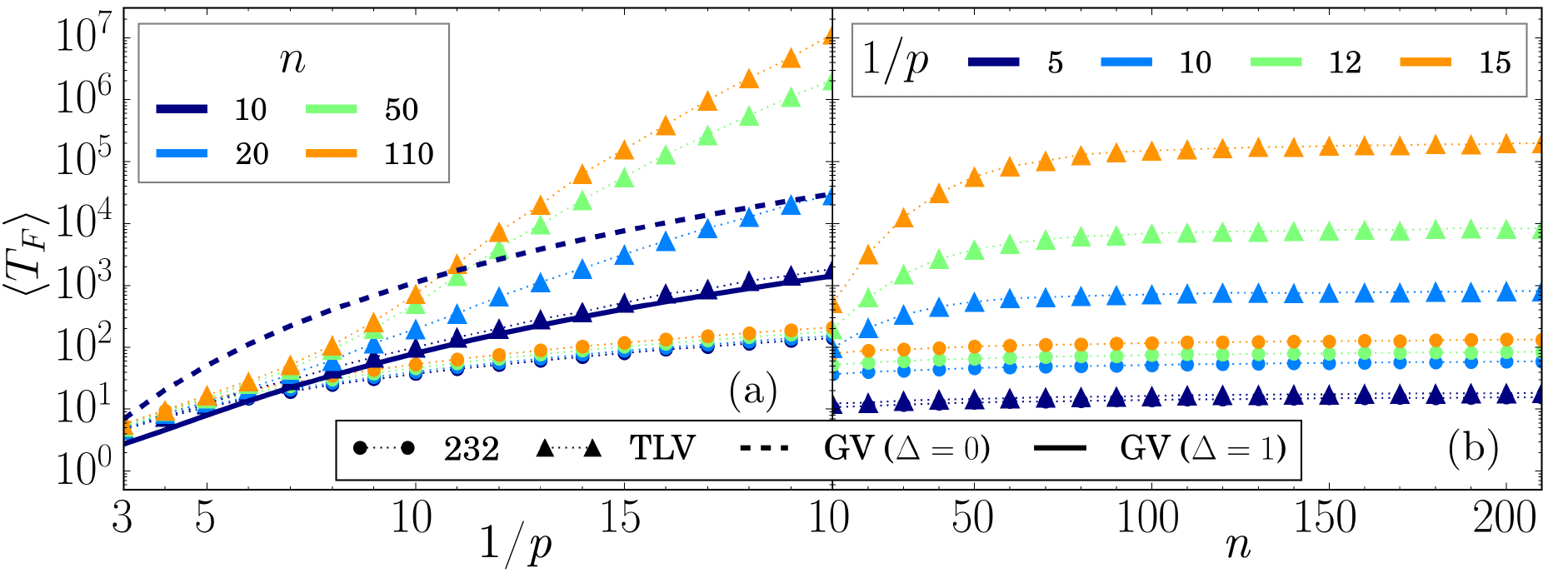}
 \caption{(a) Numerical data for flip-time (FT) dependence on the inverse bit-flip probability per cell and time step, $1/p$, for several lattice sizes $n$ (see color code). (b) FT dependence on $n$ for certain fixed $1/p$ (color coded). The 232 and TLV FTs are represented by circular and triangular data points, respectively. Analytically derived FTs for global voting~\cite{Suppl_Mat} with $n=10$ and $\Delta =0$ (dashed) or $\Delta = 2$ (solid) are also shown in (a).   }
  \label{fig2}
\end{figure}

\emph{Simulation results.}---Having quantized the 232 and TLV CAs, we now quantitatively assess their QEC capabilities. We start with the original CAs.
Fig.~\ref{fig2} shows numerical evaluations of the average number of time steps (each corresponding to one application of noise followed by the CA rule~\cite{Suppl_Mat}) necessary for the majority of cells of an all-0 initial configuration to be simultaneously found in state 1 (logical flip). We denote this quantity by flip time (FT). For each lattice size $n$ and single-cell state-flip probability $p$, 10000 orbits were sampled through Monte Carlo.  
 For comparison, we also provide corresponding plots for global voting, applying noise $1+\Delta$ times before global read-out and correction take place~\cite{Suppl_Mat}. Since time is measured in CA steps, the global-voting operational time $1+\Delta$ accounts for a possible delay $\Delta$. Fig.~\ref{fig2}(a) shows that for 10 cells global voting with $\Delta=2$ already underperforms TLV.  Recent experimental realizations of the quantum repetition code showed measurement times at least one order of magnitude larger than gate-application times~\cite{google}, so that $\Delta \gtrsim 5$ for realistic scenarios~\cite{Suppl_Mat}.

One can see from Figs.~\ref{fig2}(a),(b) that the FTs for TLV surpass the 232 values for any $n$ and $p$. The FT saturation in Fig.~\ref{fig2}(b) results from CAs approaching the infinite-lattice behavior, as expected from the wrapping lemma. Locality prevents FTs from increasing unboundedly with $n$, evidencing the absence of phase transitions~\cite{Farrelly}. Furthermore, the FTs for TLV show a characteristic inflexion at $p\approx 1/8$. Our best fitting was achieved with $\langle T_F(p,n)\rangle = 2^{c_1+\log^2_2(1/p)\big[ f_1(n) + f_2(n)\text{tanh}(c_2/p - c_3)\big]}$, with $f_i=a_i+b_i\exp{(c_in)}$ as shown in \cite{Suppl_Mat}. This function interpolates between the two regimes separated by the inflexion and approximately reproduces the lower and upper bounds found in Ref.~\cite{Park} for a similarly defined quantity. 

\begin{figure}
    \centering
    \includegraphics[width=1\linewidth]{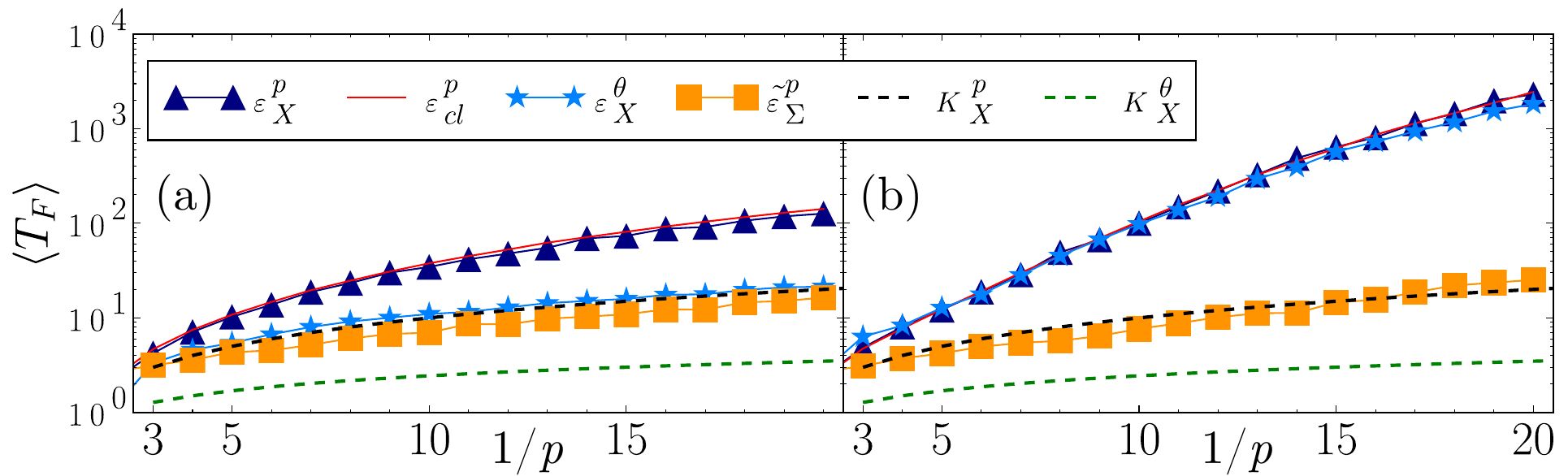}
 \caption{Flip times of Q232 (a) and QTLV (b) as functions of $1/p$. Phenomenological incoherent and coherent [with $\sin^2(\theta/2) = p$] bit-flip noise are represented by blue triangles and stars, respectively; the corresponding noise channels are represented as $\varepsilon^p_X$ and $\varepsilon^\theta_X$. Circuit incoherent depolarizing noise corresponds to orange squares, with noise channel $\tilde{\varepsilon}^p_\Sigma$. 
 For comparison, CA FTs are included as solid red curves (with channel $\varepsilon^p_{cl}$). FTs for incoherent ($K^p_X$) and coherent ($K^\theta_X$) physical-qubit noise are included as dashed black and green lines, respectively, to show the absence of a pseudothreshold (no crossings with the QCA lines). See Ref.~\cite{Suppl_Mat} for details.}
  \label{fig3}
\end{figure}

We simulate the performances of Q232 and QTLV on ProjectQ~\cite{projectq, software} via Monte Carlo sampling of $500$ random orbits of $n=12$ qubits per data point.  We use a single future (bi)string of 12 additional qubits, so that at each time step, we encode the rule update on the future register, and then decouple future and present strings by applying CNOTs controlled by the future cells. We then reset the present string to an all-0 configuration and relabel strings according to "future"$\leftrightarrow$"present". We call flip time $T_F$ the first time $t$ at which $\sum_{i=1}^n \langle \sigma^{(z)}_{i,t} \rangle < 0$. The sequence of gates is maximally parallelized: each control and target qubit is acted upon by one single gate at each circuit step, leading to circuit depths of 7 for even $n$ (cf.~\cite{Suppl_Mat}). This design allows us to run Q232 and QTLV for a 12-qubit logical state using 24 qubits. The quantum circuits for the simulation of Q232 and QTLV are shown in \cite{Suppl_Mat}. We simulate our QCAs for different physical bit-flip probabilities applied both coherently and incoherently for the phenomenological model, and incoherent depolarizing noise for the circuit-noise model, as shown in Fig.~\ref{fig3}. We start with randomly generated logical states of the form $\cos (\phi)  \otimes_i |0_{i,0}\rangle + i\sin (\phi ) \otimes_i |1_{i,0}\rangle $ with $|\phi| < \pi/4 $. We see from Fig.~\ref{fig3} that the performances of the QCAs coincide with CA ones when incoherent bit-flip noise is phenomenologically applied. 
Surprisingly, for phenomenological coherent noise, QTLV shows a considerably smaller performance drop relative to the incoherent model than the globally decoded quantum repetition code~\cite{repetition} (cf. Ref.~\cite{Suppl_Mat}).  

\emph{Concatenation.}---Both Q232 and QTLV can be concatenated to create QCAs capable of full QEC (i.e., correct both bit and phase flips). Just as the Shor code concatenates the repetition code in two complementary bases into a complete QEC code, we prove in Ref.~\cite{Suppl_Mat} that a similar design for QTLV allows for improvement in FT of a logical state afflicted by bit and phase flips. Concatenation is therefore one path towards full QEC with our proposed QCAs. 

\emph{Implementation proposal.}---Dynamically reconfigurable 2D arrays of neutral atoms allow for the 
parallelization of 2- and 3-qubit high-fidelity entangling gates~\cite{Lukin_gate, Lukin_dyn, Lukin_new, Bro, Exp_new1}. The parallel gates required for our QCAs are therefore within experimental reach~\cite{cnot_decompose, Suppl_Mat}. The reconfigurability of 2D neutral-atom arrays allows for rearrangement of time strings~\cite{Lukin_dyn, Lukin_new}. We expect Rydberg-atom platforms to be able to implement QCAs with globally applied locally conditional interactions~\cite{experiment, Lukin3} [cf. Fig.~\ref{fig1}(b)], with trapped ions offering similar capabilities~\cite{iontrap}.

\emph{Conclusion.}---In this Letter, we introduce the first QCAs for QEC. These QCAs are reversible, translation invariant, locality preserving, self-dual and independent from measurements and syndromes. They preserve coherence and information in the logical states they act upon while keeping errors local. Our simulation results highlight the robustness of QTLV, displaying remarkable performance against coherent and incoherent phenomenological noise. Our proposal for error-correcting QCAs opens the 
possibility to systematically explore and establish QCAs as a complementary paradigm for robust quantum information processing. We show that concatenation is one path towards QCA-based full QEC~\cite{Suppl_Mat}, yet higher-dimensional QCAs are also expected to have such capabilities. Topological properties of QCAs~\cite{Farrelly, index}, potentially leveraging concepts from CAs and topological QEC~\cite{Harrington, Terhal, Fibonacci, fieldca, fieldCA2}, could lead to new approaches of QCA-based measurement-free QEC.
 \\
 
\begin{acknowledgments}
 
 We thank K. Morita and P. G{\'a}cs for fruitful discussions. We gratefully acknowledge support by the ERC Starting Grant QNets through Grant No. 804247. Furthermore, MM acknowledges support by the BMBF project MUNIQC-ATOMS, the European Union’s Horizon Europe research and innovation programme under grant agreement No 101114305 (“MILLENION-SGA1” EU Project). This research is also part of the Munich Quantum Valley (K-8), which is supported by the Bavarian state government with funds from the Hightech Agenda Bayern Plus. The authors gratefully acknowledge funding by the Deutsche Forschungsgemeinschaft (DFG, German Research Foundation) through Grant No. 449905436, and under Germany’s Excellence Strategy ‘Cluster of Excellence Matter and Light for Quantum Computing (ML4Q) EXC 2004/1’ 390534769. We acknowledge computing time provided at the NHR Center NHR4CES at RWTH Aachen University (Project No. p0020074). This is funded by the Federal Ministry of Education and Research and the state governments participating on the basis of the resolutions of the GWK for national high performance computing at universities.

 \end{acknowledgments}

 \bibliography{litQCA}
 

\onecolumngrid
\clearpage

\setcounter{equation}{0}
\setcounter{figure}{0}
\setcounter{table}{0}
\setcounter{page}{1}

\makeatletter

\renewcommand{\theequation}{S\arabic{equation}}
\renewcommand{\thefigure}{S\arabic{figure}}
\renewcommand{\bibnumfmt}[1]{[S#1]}
\renewcommand{\citenumfont}[1]{S#1}
\setcounter{secnumdepth}{3}
\renewcommand\thesection{\arabic{section}}

\begin{center}
\textbf{\large \underline{Supplemental Material}}
\\[0.5cm]
\textbf{\large Quantum cellular automata for quantum error correction and density classification}
\end{center}

\begin{center}
T. L. M. Guedes, D. Winter and M. M{\"u}ller
\end{center}
 
 \section{Wolfram's nomenclature for elementary cellular automata}
 
 Elementary cellular automata are (usually periodic) 1D CAs in which each cell has 2 possible states, $(\mathbb{Z}_n, \mathbb{Z}_2,f,\{s_{i-1},s_i,s_{i+1}\})$, and the neighborhood scheme $N_d=\{s_{i-1},s_i,s_{i+1}\}$ involves the left and right neighbors of the cell the local rule $f$ acts upon, as well as the updated cell itself. Since each of the $2^3$ inputs of $f$ can be mapped into either 0 or 1 by the rule, there are in total $2^8=256$ possible choices of $f$. Wolfram developed a nomenclature for these 256 so-called elementary CA elementary-CA rules based on the decimal equivalent of the binary number obtained by listing the updates of inputs 111, 110, 101, 100, 011, 010, 001, 000, respectively~\citeS{Wolfram_statisticalS}: if $f$ gives the updates $111\to \mathbf{0}, 110\to \mathbf{0}, 101\to \mathbf{0}, 100\to \mathbf{1}, 011\to \mathbf{1}, 010\to \mathbf{1}, 001\to \mathbf{1}, 000\to \mathbf{0}$, generating an output list  $\mathbf{00011110}$, for example, that corresponds to the decimal number 30, which names the considered rule. Following this nomenclature scheme, elementary CA rules are numbered from 0 to 255.  
 
  Rule 232 corresponds to the updates $111\to \mathbf{1}, 110\to \mathbf{1}, 101\to \mathbf{1}, 100\to \mathbf{0}, 011\to \mathbf{1}, 010\to \mathbf{0}, 001\to \mathbf{0}, 000\to \mathbf{0}$. It is easy to see that its update values correspond to the majority state amongst the 3 considered cells, therefore explaining why it is also called local majority voting. 
 
 Rule 184, on the other hand, corresponds to the updates $111\to \mathbf{1}, 110\to \mathbf{0}, 101\to \mathbf{1}, 100\to \mathbf{1}, 011\to \mathbf{1}, 010\to \mathbf{0}, 001\to \mathbf{0}, 000\to \mathbf{0}$. Its update values are a slight tweak of those used by rule 232, namely the outputs of 110 and 100 are swapped relative to the latter rule. These swapped updates relative to the local majority voting are crucial to understand why rule 184 is so different: it moves ones to the right when succeeded by zero and zeros to the left when preceded by one! This These dynamics leads to the mutual annihilation of islands of zeros and ones once they meet, leaving a remnant of the largest island surrounded by alternating states. After a sufficiently long time, rule 184 leaves only islands of the same state in a sea of alternating states. Since these islands correspond to the state of the majority of the cells in the initial configuration, consecutive application of rule 232 would then allow the islands to consume the surrounding sea, resulting in perfect density classification~\citeS{FuksS}. This two-CA density classifier is, however, very ineffective in the presence of noise~\citeS{MendoncaS}; amongst other reasons, its poor performance can be exemplified through the fact that 184 moves errors around, and 0- and 1-errors are moved in opposite directions. Because of the dynamics of zeros and ones moving in opposite directions, rule 184 is also known as traffic rule. 
 
 \section{Island-growth probability under local majority voting}
 
 The probability that each cell has its state flipped at the start of a certain time step is given by $p$. For a system of $n$ cells with periodic boundary conditions, the probability that any $k<n-1$ neighboring cells flip simultaneously forming a $k$-cell island that includes a given cell $i$ is $k(1-p)^2p^k$ (here we consider that the cells bordering the island from the left and from the right are not flipped, but disregard what happens to cells further away). Those islands cannot be eroded by rule 232, and therefore after a 232 CA update only islands will be left. Once a 2-cell island forms, any new flips on its neighboring cells should increase the island size by 1 and this increase cannot be reverted by local majority voting. Furthermore, a flip in either of the island's next-neighboring cells will lead rule 232 to increase the island's size by 2, forcefully flipping the error-unafflicted cell bridging the island and the sole error. On the other hand, flipping either of the island's cells reduces its size to one, which ultimately erodes the island after a 232 update. In summary, there are 4 ways in which the 2-cell island can increase (by flipping neighbors or next-neighbors of the island) and 2 ways in which it can disappear under the action of rule 232, leading to a growth probability of 2/3 when only single flips in the vicinity of islands are considered. Consideration of multiple simultaneous flips also shows that a 2-cell island tends to grow. Applying a similar analysis to the case of 3-cell islands shows that, considering only single flips in its vicinity, the probability of island growth is 4/7 (flipping the central cell in the island does not change its size). For $k>3$, however, the island has equal probabilities of growing or shrinking when sole flips are considered. 
 
 \section{Fitting the flip time for TLV}
 
 In his thesis~\citeS{ParkS}, Park derived upper and lower bounds for a quantity he called relaxation time, $T_R$. This quantity is defined in terms of probability distributions over tuples of states (i.e., sections of the configurations in an infinite lattice). Since any finite-lattice CA is both ergodic (it has only one invariant probability distribution) and mixing (any probability distribution converges to the invariant one after sufficiently many time steps), the motivation behind the definition of the relaxation time is the investigation of the rate of information loss of a given noisy CA when only a finite fraction of an infinite lattice is investigated, as is the case for periodic finite lattices. 
 
 Denoting an $n$-cell configuration as $\zeta^{(n)}$, which is either a configuration of an $n$-cell lattice or a section of an infinite lattice, the distance between probability distributions $\mu$ and $\nu$ over such configurations can be defined as 
 \begin{equation}
     d_n (\mu , \nu ) = \sum_{\zeta^{(n)}}  \left| \mu (\zeta^{(n)}) - \nu (\zeta^{(n)} ) \right| .
 \end{equation}
 Park then defines the relaxation time as 
 \begin{equation}
     T_R(n, \rho, \text{CA}') = \text{min} \{ t \;|\; \text{sup}_{\mu, \nu} \, d_n (F'^{(t)}\mu , F'^{(t)}\nu ) <\rho \},
 \end{equation}
 where $\text{sup}$ stands for supremum and the prime on the compactified notation "CA" (which includes the lattice, states per cell, neighborhood scheme and transition function) and on its (composed) global rule $F'^{(t)}$ means that noise is applied with probability $p$ after every CA step. $F'^{(t)}\mu$ then denotes the probability distribution of configurations resulting from the application of the CA's (noisy) global rule $t$ times on the configurations of $\mu$.  Park showed that for sufficiently small $p$ the relaxation time has values in the range
 \begin{equation}
     2^{c_1\log^2 (1/p)} < T_R (n, \rho, \text{CA}' ) < 2^{c_2\log^2 (1/p)} 
 \end{equation}
 for some $c_1$ and $c_2$. Note that the logarithm basis does not matter, since a basis-change factor could be absorbed into $c_1$ and $c_2$.
 
\begin{figure}
    \centering
    \includegraphics[width=\textwidth]{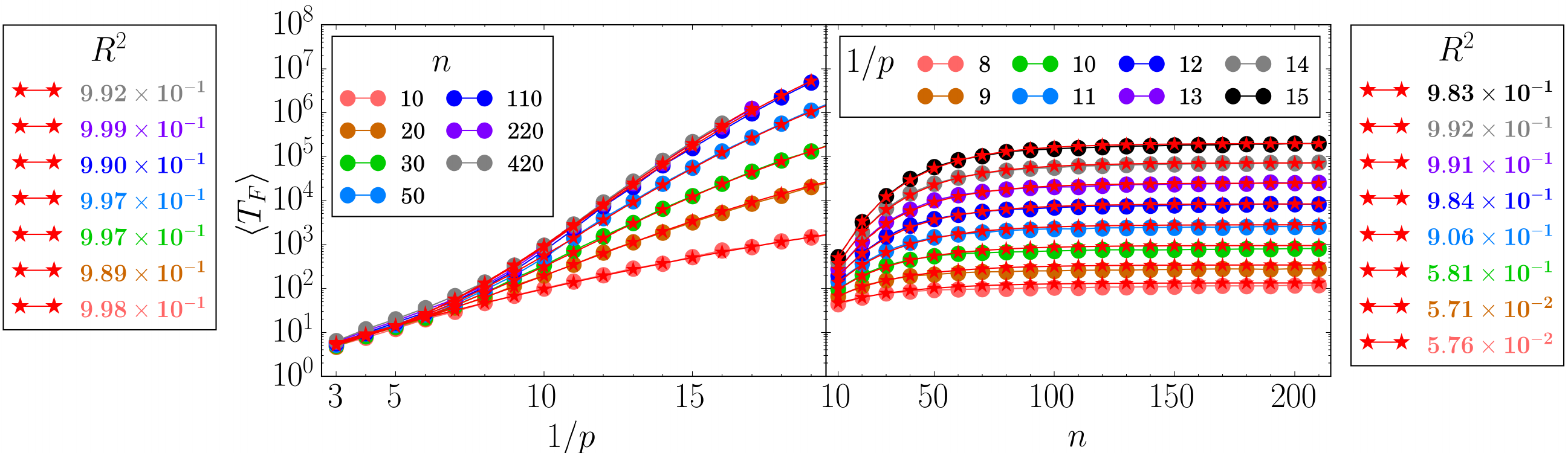}
    \caption{TLV flip times as a function of the inverse noise $1/p$ (left) and of the lattice size $n$ (right). The different color-coded curves correspond to several values of $n$ ($1/p$) on the left (right) panel. The corresponding fitting curves generated from Eq.~\eqref{flip_fit} are shown in red, and the least-squares coefficients $R^2$ for each of these curves is shown on the side of the corresponding panel.}
    \label{fig_flip_fit}
\end{figure}

 Using Park's bounds as a reference, we investigated several functions that interpolate between regimes described by $k 2^{c(p,n)\log^2 (1/p)}$ for different choices of $k$ and $c(p,n)$. Those functions were fitted on the data collected for the flip times of our CAs, a more practical quantity than the mathematically rigorously defined relaxation time. It is worth noting that those two quantities are qualitatively connected, since both measure the memory time of the investigated CAs. Our fitting results show that this connection is more than simply qualitative, with the best achieved flip-time fitting function for TLV being given by
\begin{equation}
   \langle T_F(p, n)\rangle  = 2^{1.5+\{0.71 - 0.36\exp(-0.036n)+[0.53-0.72\exp(-0.04n)]\text{tanh}[0.136(1/p-9.3)]\}\log_2(1/p)}.
    \label{flip_fit}
\end{equation}
The comparison between Eq.~\eqref{flip_fit} and the numerically generated data for TLV is shown in Fig.~\ref{fig_flip_fit}, with color-coded least-squares coefficients $R^2$ given on the right for several $n$ values and on the left for several $1/p$ values. Note that, since Park's bounds hold for sufficiently low $p$, our fitting function derived from those bounds shows decreasing $R^2$ values as $p$ increases. 
    
 \section{Global majority voting}
 
 We address here the classical counterpart of the repetition code. In a 1D lattice $L^g_1$ of $n$ two-state cells (bits), we consider that, at the end of a period of $1+\Delta$ (discretized) time steps, the configuration of the system will be given by the state in which the majority of cells is found, i.e., $\zeta^g(j) \to \lfloor 1/2 + \sum_{i=0}^{n-1} s^g_i/n\rfloor \; \forall \, j\in L^g_1$ ($\lfloor \cdot \rfloor$ denotes the floor/round-down operation). We use the superscript $g$ to highlight that the quantities we address here are mere translations of the respective CA quantities, but actually refer to a globally dependent discrete-time update process. Describing the two states of each cell as $s^g_i\in \{ 0, 1\}$, we assume that the noise acts (independently) on each cell with a probability $p$ per time step, flipping the state of this cell before the next global evaluation and update take place. 
 
 Starting from an all-0 configuration [i.e., $\zeta^g(i)=0 \; \forall \, i \in L^g_1$], the state of each cell after $t\leq 1+\Delta$ time steps will have been flipped by noise with a probability determined by all sequences of elementary events in which an overall odd number of flips happened: 
 \begin{equation}
     p(t)= \sum_{k\in \{ 0, \ldots , t\} | k \; \text{odd}} \binom{t}{k} (1-p)^{t-k}p^k = \frac{1}{2}[(1-p)+p]^t - \frac{1}{2}[(1-p)-p]^t= \frac{1}{2}[1-(1-2p)^t].
 \end{equation}
 The probability that the state of a cell has been flipped right before an update is therefore given by $p(1+\Delta)$. The update at time $1+\Delta$ will force the system into the all-1 configuration if more than half of the cells have been flipped, what happens with a probability
 \begin{equation}
     P(t)=\sum_{j\geq \lceil n/2 \rceil} \binom{n}{j}[1-p(t)]^{n-j} p^j(t)
     .
     \label{logical_P_maj}
 \end{equation}
 From these relations it becomes clear that the logical-flip probability per update is given by $P(1+\Delta)$. 
 
 We can describe the dynamics of the entire system in terms of a single logical bit that gets updated at time steps of duration $1+\Delta$. The update takes the logical state to itself with probability $1-P(1+\Delta)$ or to its complement with probability $P(1+\Delta)$. The probability that the logical flip happens at the $k$-th update step is given by a geometric distribution, $P_k(1+\Delta) = [1-P(1+\Delta)]^{k-1}P(1+\Delta)$. The average time taken for a logical flip to happen is therefore given by
 \begin{equation}
     \langle T_F \rangle = \sum_{k=0}^\infty k P_k(1+\Delta) = P(1+\Delta) \frac{\partial }{\partial \bar{P} } \sum_{k=0}^\infty \bar{P}^k  = P(1+\Delta) \frac{\partial }{\partial \bar{P}  } \frac{1}{1-\bar{P}  }= \frac{1}{P(1+\Delta)},
     \label{inverse_relation}
 \end{equation}
 where $\bar{P} =1-P(1+\Delta)$ is temporarily assumed to be a continuous variable. It is therefore clear that the average flip time of a system globally updated is the inverse of the logical-flip probability. This is a general feature of the geometric distribution. 
 
\begin{figure}[h!]
    \centering
    \includegraphics[width=\textwidth]{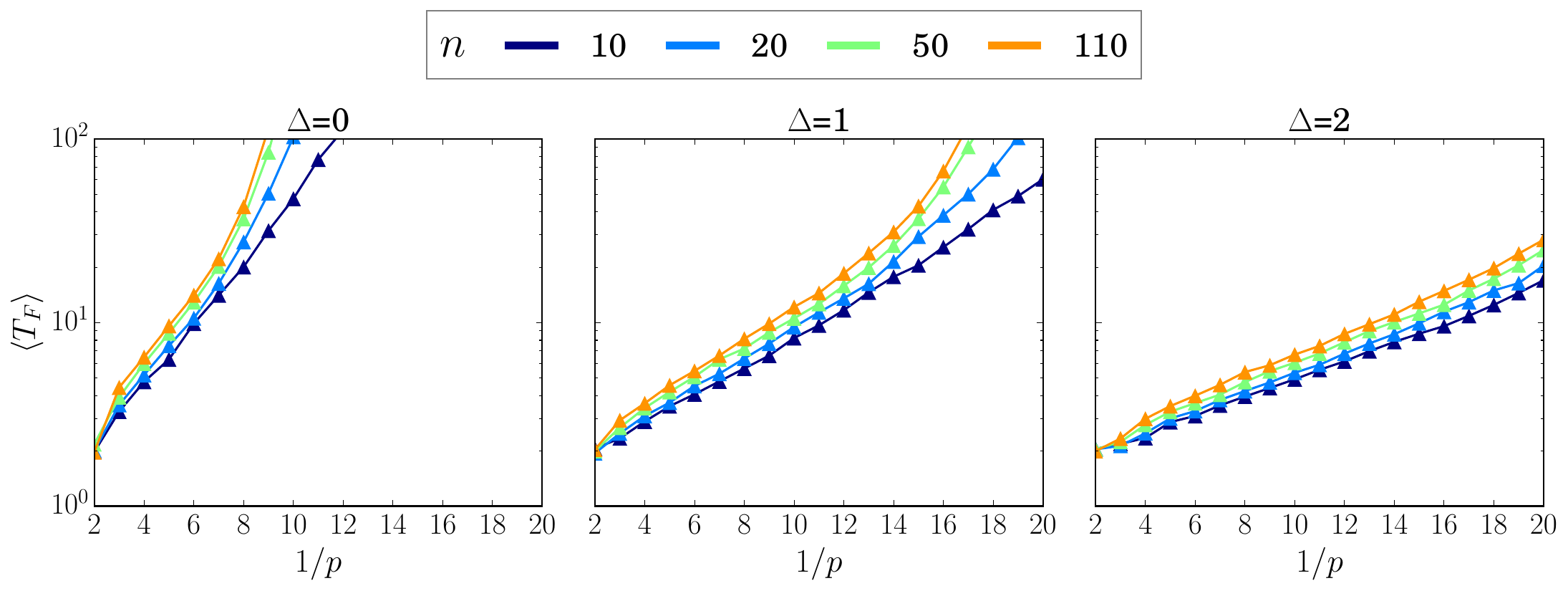}
    \caption{Flip times for the quantum repetition code under phenomenological incoherent noise afflicting both data and ancilla qubits with bit-flip probability $p$. Syndromes are collected at intervals of $1+\Delta$ QCA time steps, and the flip time corresponds to the minimal number $T$ of syndrome-collection rounds that allows for proper decoding of the logical state when using minimum-weight perfect matching. In this scenario, the state is only truly corrected classically, i.e., after all data qubits have been measured. Notice that measuring the data qubits is needed to guarantee that the logical state is projected back into the code space: by relying solely on faulty syndrome information, the logical state will most probably be outside of the code space, as is the case when purely spatial minimum-weight perfect matching is employed.}
  \label{MWPM}
\end{figure}

 It is important to clarify that the update time for global majority voting  was chosen as $1+\Delta$ because the units of time correspond to standard CA time steps. Since, as in the quantum case, global updates require observation of the states of all (dual) cells in the lattice, it is reasonable to assume that one or more CA updates take place in the same time interval taken by the global majority voting to perform one update. For this reason, the quantity $\Delta$ is called delay. In fact, for some platforms the measurement times might even scale with the system size, $1+\Delta=\lceil n/c \rceil$ for some measurement speed $c$ on $n$ qubits~\citeS{BuechlerS}. In a recent state-of-the-art experimental implementation of the quantum repetition code in a set of up to 21 qubits (11 physical data qubits and 10 ancillas), the recorded parallel measurement and reset time was 880 nanoseconds~\citeS{googleS}. In comparison, in the same experimental setup a depth-4 circuit was implement in 80 ns, giving an average of 20 ns per unit of circuit depth. Our QCAs can be implemented as depth-7 circuits, which would correspond to $\sim 140$ ns per QCA step, a value about 6 times smaller than the measurement and reset time registered in the experiment. From this comparison, we assume that state-of-the-art implementations of the (quantum) repetition code perform updates at periods of $1+\Delta \gtrsim 6$, while the QCAs take a single time step per update. It is worth noting, however, that in the quantum case the global update might fail to bring the logical state back to the computational basis of (superpositions of) all-0 and all-1 configurations, since syndrome-based corrections are not implemented in all qubits, but only on specific qubits that are believed to have been flipped by noise, which could result in a logical state with a few flipped qubits (e.g., when the syndrome collection and/or processing is faulty). In this sense, the (classical) global majority voting gives an upper bound for the flip time of the quantum repetition code. In fact, if one implements the quantum repetition code with noisy ancillas and considers the application of minimum-weight perfect matching (MWPM) on syndrome data distributed in space-time (i.e., along the lattice formed by the time-$T$ slice of history of the 1D system) when measurements are perfect, the smallest possible $T$ time steps at which the logical information is incorrectly decoded (i.e., a logical flip is enforced by space-time-global MWPM) can be considered the repetition-code counterpart of the flip time. The corresponding quantity is plotted for different lattice sizes, physical bit-flip probabilities (under phenomenological incoherent noise) and delays in Fig.~\ref{MWPM}. Application of MWPM in space-time syndrome data requires collection of data globally in space and time, therefore a single classical correction step is applied on the read-out data after $T$ time steps, once the data qubits have been measured and the resulting space-time data has been processed. If delays are considered, the collection of syndromes will take place in $1+\Delta$ steps, therefore phenomenological incoherent bit-flip noise is applied $(1+\Delta)T$ times on the system before correction takes place. The results in Fig.~\ref{MWPM} show that flip times in the quantum repetition code are reduced drastically as the delay increases, and at delays $\Delta \geq 2$ there is little gain in flip time when increasing the lattice size for $1/2 \geq p \geq 1/20$. When compared with QTLV for 12 qubits, the quantum repetition code seems to underperform for all investigated lattice sizes and bit-flip probabilities whenever the delay is non-zero; however, a direct comparison is unreasonable due to the fact that our QCA simulations consider no equivalent of the ancilla-qubit noise, as discussed below. 

 Lastly, we reiterate that Eq.~\eqref{inverse_relation} allows us to convert flip times into logical-flip probabilities and vice versa whenever the logical-flip probabilities are described by a geometric distribution. We can therefore also estimate flip times for the quantum repetition code simulated under different noise scenarios from the logical-flip-probability data provided in Ref.~\citeS{repetitionS}. Although the repetition code only really obeys a geometric distribution if the system ends in the code space after each error-correction cycle, so that flip times would be measured in terms of QEC cycles, we can make rough estimates using Eq.~\eqref{inverse_relation} also when a single extended QEC cycle with space-time decoding is considered. For a phenomenological noise model with $p=1/12$ (also for ancilla qubits) and $n=11$, Ref.~\citeS{repetitionS} gives $P(1) \approx 0.05$ and $P(1) \approx 0.1$ when incoherent and coherent noise are considered, respectively. This would correspond to $\langle T_F \rangle \sim 20$ and $\langle T_F \rangle\sim 10$ for the incoherent and coherent cases. The high values of $P(1)$ for the simulated repetition code with global decoding, even with $\Delta =0$, can be attributed to the application of noise on the ancilla qubits, which can lead to faulty syndrome estimation and therefore deteriorate the correction performance. In fact, a pair of false syndromes can lead to correction-induced errors in an entire set of data qubits and therefore have a much larger weight on the performance deterioration of the repetition code than data-qubit errors. A direct comparison with the Q232 and QTLV flip times is not possible because we do not consider noise acting on future cells (the closest counterpart of ancilla noise); nonetheless, this scenario is approximately reproduced by considering $p(2)$ instead of $p$ as the physical bit-flip probability, since each QCA step (besides the first and last ones) would be both preceded and succeeded by the action of noise with probability $p$, i.e., noise would act in both the present and future registers at each time step. When $p=1/12$, $p(2)=11/72 \approx 1/7$ and the corresponding flip times obtained for the QTLV with $n=12$ are $\langle T_F \rangle =28.8$ (incoherent) and $\langle T_F \rangle =28.3$ (coherent). In comparison, the results shown in Fig.~\ref{MWPM} for $n=10$ give $\langle T_F \rangle =112$ (incoherent, $\Delta = 0$), $\langle T_F \rangle =11.7$ (incoherent, $\Delta = 1$) and $\langle T_F \rangle =6.1$ (incoherent, $\Delta = 2$). We are therefore led to conclude that the QTLV is a competitive error-correcting architecture even when compared to the repetition code.

\section{Toffoli's approach to convert irreversible CAs into reversible ones}

In Ref.~\citeS{ToffoliCAS}, Toffoli introduced a scheme to convert irreversible CAs into reversible ones. This scheme requires extending $L_d$ to $L_{d+1}$, where the additional dimension fulfils the role of a discrete-time counter. In fact, with each time step, the register in $L_d$ is fed with a new register with cells set to the all-0 configuration, so that the CA rule can map information from one register to the other. If we set the states of the cells $(i_1,t), (i_2,t), (i_3,t), (i_4,t+1)$ to $s_{i_1, t}=s_{1}$, $s_{i_2, t}=s_{2}$, $s_{i_3, t}=s_{3}$, $s_{i_4, t+1}=s_{4}$ and represent the complements (bit flips) of those values by $\overline{s_1}$, $\overline{s_2}$, $\overline{s_3}$, $\overline{s_4}$, the truth table for the extended rule reads
\begin{table}[h!]
	\centering
	\begin{tabular}{l c c c|c c c r}
	\multicolumn{4}{c}{Input} & \multicolumn{4}{c}{Output}\\
	\hline
	$s_1$ & $s_2$ & $s_3$ & $s_4$ & $s_1$ & $s_2$ & $s_3$ & $[s_4 + f(s_1, s_2, s_3)]$ \\
	$s_1$ & $s_2$ & $\overline{s_3}$ & $s_4$ & $s_1$ & $s_2$ & $\overline{s_3}$ & $[s_4 + f(s_1, s_2, \overline{s_3})]$ \\
	$s_1$ & $\overline{s_2}$ & $s_3$ & $s_4$ & $s_1$ & $\overline{s_2}$ & $s_3$ & $[s_4 + f(s_1, \overline{s_2}, s_3)]$ \\
	$s_1$ & $\overline{s_2}$ & $\overline{s_3}$ & $s_4$ & $s_1$ & $\overline{s_2}$ & $\overline{s_3}$ & $[s_4 + f(s_1, \overline{s_2}, \overline{s_3})]$ \\
	$\overline{s_1}$ & $s_2$ & $s_3$ & $s_4$ & $\overline{s_1}$ & $s_2$ & $s_3$ & $[s_4 + f(\overline{s_1}, s_2, s_3)]$ \\
	$\overline{s_1}$ & $s_2$ & $\overline{s_3}$ & $s_4$ & $\overline{s_1}$ & $s_2$ & $\overline{s_3}$ & $[s_4 + f(\overline{s_1}, s_2, \overline{s_3})]$ \\
	$\overline{s_1}$ & $\overline{s_2}$ & $s_3$ & $s_4$ & $\overline{s_1}$ & $\overline{s_2}$ & $s_3$ & $[s_4 + f(\overline{s_1}, \overline{s_2}, s_3)]$ \\
	$\overline{s_1}$ & $\overline{s_2}$ & $\overline{s_3}$ & $s_4$ & $\overline{s_1}$ & $\overline{s_2}$ & $\overline{s_3}$ & $[s_4 + f(\overline{s_1}, \overline{s_2}, \overline{s_3})]$ \\
	\hline
	$s_1$ & $s_2$ & $s_3$ & $\overline{s_4}$ & $s_1$ & $s_2$ & $s_3$ & $[\overline{s_4} + f(s_1, s_2, s_3)]$ \\
	$s_1$ & $s_2$ & $\overline{s_3}$ & $\overline{s_4}$ & $s_1$ & $s_2$ & $\overline{s_3}$ & $[\overline{s_4} + f(s_1, s_2, \overline{s_3})]$ \\
	$s_1$ & $\overline{s_2}$ & $s_3$ & $\overline{s_4}$ & $s_1$ & $\overline{s_2}$ & $s_3$ & $[\overline{s_4} + f(s_1, \overline{s_2}, s_3)]$ \\
	$s_1$ & $\overline{s_2}$ & $\overline{s_3}$ & $\overline{s_4}$ & $s_1$ & $\overline{s_2}$ & $\overline{s_3}$ & $[\overline{s_4} + f(s_1, \overline{s_2}, \overline{s_3})]$ \\
	$\overline{s_1}$ & $s_2$ & $s_3$ & $\overline{s_4}$ & $\overline{s_1}$ & $s_2$ & $s_3$ & $[\overline{s_4} + f(\overline{s_1}, s_2, s_3)]$ \\
	$\overline{s_1}$ & $s_2$ & $\overline{s_3}$ & $\overline{s_4}$ & $\overline{s_1}$ & $s_2$ & $\overline{s_3}$ & $[\overline{s_4} + f(\overline{s_1}, s_2, \overline{s_3})]$ \\
	$\overline{s_1}$ & $\overline{s_2}$ & $s_3$ & $\overline{s_4}$ & $\overline{s_1}$ & $\overline{s_2}$ & $s_3$ & $[\overline{s_4} + f(\overline{s_1}, \overline{s_2}, s_3)]$ \\
	$\overline{s_1}$ & $\overline{s_2}$ & $\overline{s_3}$ & $\overline{s_4}$ & $\overline{s_1}$ & $\overline{s_2}$ & $\overline{s_3}$ & $[\overline{s_4} + f(\overline{s_1}, \overline{s_2}, \overline{s_3})]$
	\end{tabular}
\end{table}

Note that, since the cell states are binary, $[\overline{s_4} + f(\cdot, \cdot, \cdot)]= \overline{[{s_4} + f(\cdot, \cdot, \cdot)]}$.
It is clear that the entries in this table, when expressed in terms of binary numbers of the form $s_1 s_2 s_3 s_4$, correspond to a $16\times 16$ invertible matrix. This invertibility renders the extended CA rule reversible. It is worth pointing out that $s_{i_4,t+1}\to s_{i_2,t+1}$ for the 232 CA, while $s_{i_4,t+1}$ is a completely independent cell in the TLV CA. 

\section{QCA properties of Q232 and QTLV}

As discussed in Ref.~\citeS{SchumacherS}, the locality-preserving character of a QCA is determined by the image of the automorphism $u$ when the domain is restricted to a local algebra $\mathcal{A}_i$. That image consists of observables that act nontrivially on at most all cells in the region $\mathcal{R}_i\subset \Gamma_d$. In other words, the effects of local observables cannot be propagated by the QCA beyond $\mathcal{R}_i$ within a single discrete time step.  For this reason, this property is also commonly referred to as causality, since cells arbitrarily far from each other cannot be correlated by observables in a single time step, similarly to how spacelike-separated observers in relativity cannot signal to each other. 
This definition of $\mathcal{R}_i$, however, is somewhat abstract, and for this reason a more intuitive approach to find this region is also presented in Ref.~\citeS{SchumacherS}.

The protocol to find $\mathcal{R}_i$ is based on the application of quasi-commuting quasi-local unitaries $U_j$ for $j$ running through a region $\Delta_i$ of increasingly larger extension in $\Gamma_d$ and centered at the cell $i$ of the considered local algebra. This protocol is expressed compactly as 
\begin{equation}
    u(a_i) = \lim_{\Delta_i \to \Gamma_d} U^\dag_{\Delta_i} a_i U_{\Delta_i}
\end{equation}
with $U_{\Delta_i}= \prod_{j\in \Delta_i} U_j$. It is worth noting that for a locality-preserving QCA derived from a CA this limit is reached once $\Delta_i = N_d$, and therefore further increasing the extension of $\Delta_i$ does not affect $u(a_i)$. In other words, one only needs a limited number of operators $U_i$ to implement the automorphism on each cell, even if the lattice is infinite.  

We consider the su(2) algebra of observables. In fact, $u$ being a homomorphism means that we only need to derive $u(\sigma^{(x)}_i)$ and $u(\sigma^{(y)}_i)$, since $u(\mathbb{1})=\mathbb{1}$, $u(\sigma^{(z)}_i)=-iu(\sigma^{(x)}_i)u(\sigma^{(y)}_i)$ and $u(\sum_{m,i} c_{m,i}\sigma^{(m)}_i)=\sum_{m,i} c_{m,i} u(\sigma^{(m)}_i)$. 

For Q232 it is possible to show that $U_{N_d,i,t}=U_{i+1,t}U_{i,t}U_{i-1,t}$ suffices to evolve $\sigma^{(m)}_{i,t}$. For the sake of visualization, we represent spatiotemporally distributed (quasi-)local observables according to
\begin{equation}
 a_{i,t}=
    \begin{Bmatrix}
 & \mathbb{1}_{i-1,t+1} & \mathbb{1}_{i,t+1} & \mathbb{1}_{i+1,t+1} & \\
\mathbb{1}_{i-2,t} & \mathbb{1}_{i-1,t} & a_{i,t} & \mathbb{1}_{i+1,t} & \mathbb{1}_{i+2,t} \\
& & \mathbb{1}_{i,t-1} & &
\end{Bmatrix} = 
    \begin{Bmatrix}
 & \mathbb{1} & \mathbb{1} & \mathbb{1} &  \\
\mathbb{1} & \mathbb{1} & a & \mathbb{1} & \mathbb{1} \\
& & \mathbb{1} & &
\end{Bmatrix},
\label{space_rep}
\end{equation}
where the local observables in the entries within the braces are multiplied to form a single quasi-local observable. Empty entries within the braces are always identities. Note that the top, center and bottom rows correspond to times $t+1$, $t$ and $t-1$, respectively. The central column corresponds to site $i$, while columns to its right or left are associated with cells located to the right or to the left of $i$. Representation~\eqref{space_rep} allows us to
reproduce the geometrical arrangement of the cells the considered observables act on. Accordingly, the evolution of $\sigma^{(x)}_{i,t}$ can be represented as
\begin{eqnarray}
    U_{N_d,i,t}^\dag \sigma^{(x)}_{i,t} U_{N_d,i,t}= 
    \begin{Bmatrix}
 & \mathbb{1} & \mathbb{1} & \mathbb{1} &  \\
\pi_+ & \pi_+ & \sigma^{(x)} & \pi_+ & \pi_+ \\
 & & \sigma^{(x)} & & 
\end{Bmatrix}
+
    \begin{Bmatrix}
 & \mathbb{1} & \mathbb{1} & \sigma^{(x)} &  \\
\pi_+ & \pi_+ & \sigma^{(x)} & \pi_+ & \pi_- \\
 & & \sigma^{(x)} & & 
\end{Bmatrix}
+    \begin{Bmatrix}
 & \mathbb{1} & \sigma^{(x)} & \sigma^{(x)} &  \\
\pi_+ & \pi_+ & \sigma^{(x)} & \pi_- & \pi_+ \\
 & & \sigma^{(x)} & & 
\end{Bmatrix}
+ \nonumber\\
    \begin{Bmatrix}
 & \sigma^{(x)} & \sigma^{(x)} & \mathbb{1} &  \\
\pi_+ & \pi_- & \sigma^{(x)} & \pi_+ & \pi_+ \\
 & & \sigma^{(x)} & & 
\end{Bmatrix}
+
    \begin{Bmatrix}
 & \sigma^{(x)} & \mathbb{1} & \mathbb{1} &  \\
\pi_- & \pi_+ & \sigma^{(x)} & \pi_+ & \pi_+ \\
 & & \sigma^{(x)} & & 
\end{Bmatrix}
+    \begin{Bmatrix}
 & \mathbb{1} & \sigma^{(x)} & \mathbb{1} &  \\
\pi_+ & \pi_+ & \sigma^{(x)} & \pi_- & \pi_- \\
 & & \sigma^{(x)} & & 
\end{Bmatrix}
+ 
 \begin{Bmatrix}
 & \sigma^{(x)} & \sigma^{(x)} & \sigma^{(x)} &  \\
\pi_+ & \pi_- & \sigma^{(x)} & \pi_+ & \pi_- \\
 & & \sigma^{(x)} & & 
\end{Bmatrix}
+\nonumber \\
    \begin{Bmatrix}
 & \sigma^{(x)} & \mathbb{1} & \sigma^{(x)} &  \\
\pi_- & \pi_+ & \sigma^{(x)} & \pi_+ & \pi_- \\
 & & \sigma^{(x)} & & 
\end{Bmatrix}
+
    \begin{Bmatrix}
 & \sigma^{(x)} & \sigma^{(x)} & \sigma^{(x)} &  \\
\pi_- & \pi_+ & \sigma^{(x)} & \pi_- & \pi_+ \\
 & & \sigma^{(x)} & & 
\end{Bmatrix}
+    \begin{Bmatrix}
 & \mathbb{1} & \sigma^{(x)} & \mathbb{1} &  \\
\pi_- & \pi_- & \sigma^{(x)} & \pi_+ & \pi_+ \\
 & & \sigma^{(x)} & & 
\end{Bmatrix}
+  \begin{Bmatrix}
 & \sigma^{(x)} & \mathbb{1} & \sigma^{(x)} &  \\
\pi_+ & \pi_- & \sigma^{(x)} & \pi_- & \pi_+ \\
 & & \sigma^{(x)} & & 
\end{Bmatrix}
+\nonumber\\
    \begin{Bmatrix}
 & \mathbb{1} & \mathbb{1} & \sigma^{(x)} &  \\
\pi_- & \pi_- & \sigma^{(x)} & \pi_- & \pi_+ \\
 & & \sigma^{(x)} & & 
\end{Bmatrix}
+
    \begin{Bmatrix}
 & \mathbb{1} & \sigma^{(x)}  & \sigma^{(x)} &  \\
\pi_- & \pi_- & \sigma^{(x)} & \pi_+ & \pi_- \\
 & & \sigma^{(x)} & & 
\end{Bmatrix}
+    \begin{Bmatrix}
 & \sigma^{(x)} & \sigma^{(x)} & \mathbb{1} &  \\
\pi_- & \pi_+ & \sigma^{(x)} & \pi_- & \pi_- \\
 & & \sigma^{(x)} & & 
\end{Bmatrix}
+  \begin{Bmatrix}
 & \sigma^{(x)} & \mathbb{1} & \mathbb{1} &  \\
\pi_+ & \pi_- & \sigma^{(x)} & \pi_- & \pi_- \\
 & & \sigma^{(x)} & & 
\end{Bmatrix}
+ \nonumber\\
    \begin{Bmatrix}
 & \mathbb{1}  & \mathbb{1} & \mathbb{1} &   \\
\pi_- & \pi_- & \sigma^{(x)} & \pi_- & \pi_- \\
 & & \sigma^{(x)} & & 
\end{Bmatrix},
\hspace{9.2 cm} 
\label{toolong}
\end{eqnarray}
with projectors $\pi_\pm =\frac{1}{2}[\mathbb{1}\pm \sigma^{(z)}]$.
As an example, the first term on the right-hand side of Eq.~\eqref{toolong} represents the action of a bit-flip on the time-$t$ central cell when the present register has a sequence of states $|0_{i-2,t}0_{i-1,t}q_{i,t}0_{i+1,t}0_{i+2,t}\rangle$ around cell $i$. Regardless of the value of $q_{i,t}$, a physical bit-flip on this cell will not affect the Q232 outcome, therefore the time-$(t+1)$ cells are unaffected or, equivalently, acted upon by the identity. Flipping the state $q_{i,t}$, however, also changes the decoupling outcome at the $(t-1)$-register, therefore the bit-flip operator propagates into the past. Thanks to self-duality, this "backward bit-flip propagation" has no effect on the decoupling, since it equally affects both quantum configurations in superposition. Inspection of Eq.~\eqref{toolong} shows that a single bit-flip can propagate to at most 3 cells in the future register.

For the remaining local observable we have
\begin{equation}
     U_{N_d,i,t}^\dag \sigma^{(z)}_{i,t} U_{N_d,i,t}= \sigma^{(z)}_{i,t} .
\end{equation}
Phase flips are therefore not propagated in any direction by our QCAs.

Since we are dealing with QCAs defined on finite lattices, it is possible to define a global unitary that performs the automorphism by simply multiplying all quasi-local unitaries. This leads to the global Q232 unitary
\begin{equation}
   U_{t}^{232}=  \exp \left\{\pm i\pi\sum_i\left[  b_{i,t-1}{c_{i,t}}+ b_{i,t+1}{(c_{i+1,t}c_{i-1,t}+c_{i+1,t}c_{i,t}+c_{i,t}c_{i-1,t})}\right]\right\}
\end{equation}
with $b_{i,t}=[\mathbb{1}_{i,t}-\sigma^{(x)}_{i,t}]/2$ and $c_{i,t}=[\mathbb{1}_{i,t}-\sigma^{(z)}_{i,t}]/2$.

The derivations pertaining QTLV are rather similar. One has, however, two rules, one for each value of $j$. For the action of the $(j=+1)$-automorphism on a local observable algebra, we get
\begin{eqnarray}
   ( U^{(+1)}_{N_d,i,t})^\dag \sigma^{(x,+1)}_{i,t} U^{(+1)}_{N_d,i,t}= \hspace{11cm}\nonumber \\
    \begin{Bmatrix}
 &  & \mathbb{1} & \mathbb{1}  \\
  & & & \\
\pi_+ & \sigma^{(x)} & \pi_+  &  \\
 & &  \pi_+ & \pi_+ \\
 & \sigma^{(x)} & & 
\end{Bmatrix}
+
    \begin{Bmatrix}
 &  & \mathbb{1} & \mathbb{1}  \\
  & & & \\
\pi_+ & \sigma^{(x)} & \pi_-  &  \\
 & &  \pi_+ & \pi_- \\
 & \sigma^{(x)} & & 
\end{Bmatrix}
+    \begin{Bmatrix}
 &  & \mathbb{1} & \sigma^{(x)}  \\
  & & & \\
\pi_+ & \sigma^{(x)} & \pi_+  &  \\
 & &  \pi_+ & \pi_- \\
 & \sigma^{(x)} & & 
\end{Bmatrix}
+ 
    \begin{Bmatrix}
 &  & \mathbb{1} & \sigma^{(x)}  \\
  & & & \\
\pi_+ & \sigma^{(x)} & \pi_-  &  \\
 & &  \pi_+ & \pi_+ \\
 & \sigma^{(x)} & & 
\end{Bmatrix}
+\nonumber\\
    \begin{Bmatrix}
 &  & \mathbb{1} & \mathbb{1}  \\
  & & & \\
\pi_+ & \sigma^{(x)} & \pi_+  &  \\
 & &  \pi_- & \pi_+ \\
 & \sigma^{(x)} & & 
\end{Bmatrix}
+   \begin{Bmatrix}
 &  & \mathbb{1} & \mathbb{1}  \\
  & & & \\
\pi_- & \sigma^{(x)} & \pi_-  &  \\
 & &  \pi_- & \pi_- \\
 & \sigma^{(x)} & & 
\end{Bmatrix}
+ 
    \begin{Bmatrix}
 &  & \mathbb{1} & \sigma^{(x)}  \\
  & & & \\
\pi_- & \sigma^{(x)} & \pi_+  &  \\
 & &  \pi_- & \pi_- \\
 & \sigma^{(x)} & & 
\end{Bmatrix}
+ 
    \begin{Bmatrix}
 &  & \mathbb{1} & \sigma^{(x)}  \\
  & & & \\
\pi_- & \sigma^{(x)} & \pi_-  &  \\
 & &  \pi_- & \pi_+ \\
 & \sigma^{(x)} & & 
\end{Bmatrix}
+\nonumber \\
   \begin{Bmatrix}
 &  & \sigma^{(x)} & \mathbb{1}  \\
  & & & \\
\pi_- & \sigma^{(x)} & \pi_+  &  \\
 & &  \pi_+ & \pi_+ \\
 & \sigma^{(x)} & & 
\end{Bmatrix}
+   \begin{Bmatrix}
 &  & \sigma^{(x)} & \mathbb{1}  \\
  & & & \\
\pi_- & \sigma^{(x)} & \pi_-  &  \\
 & &  \pi_+ & \pi_- \\
 & \sigma^{(x)} & & 
\end{Bmatrix}
+  \begin{Bmatrix}
 &  & \sigma^{(x)} & \sigma^{(x)}  \\
  & & & \\
\pi_- & \sigma^{(x)} & \pi_+  &  \\
 & &  \pi_+ & \pi_- \\
 & \sigma^{(x)} & & 
\end{Bmatrix}
+
   \begin{Bmatrix}
 &  & \sigma^{(x)} & \sigma^{(x)}  \\
  & & & \\
\pi_- & \sigma^{(x)} & \pi_-  &  \\
 & &  \pi_+ & \pi_+ \\
 & \sigma^{(x)} & & 
\end{Bmatrix}
+\nonumber\\
   \begin{Bmatrix}
 &  & \sigma^{(x)} & \mathbb{1}  \\
  & & & \\
\pi_+ & \sigma^{(x)} & \pi_+  &  \\
 & &  \pi_- & \pi_+ \\
 & \sigma^{(x)} & & 
\end{Bmatrix}
+      \begin{Bmatrix}
 &  & \sigma^{(x)} & \mathbb{1}  \\
  & & & \\
\pi_+ & \sigma^{(x)} & \pi_-  &  \\
 & &  \pi_- & \pi_- \\
 & \sigma^{(x)} & & 
\end{Bmatrix}
+    \begin{Bmatrix}
 &  & \sigma^{(x)} & \sigma^{(x)}  \\
  & & & \\
\pi_+ & \sigma^{(x)} & \pi_+  &  \\
 & &  \pi_- & \pi_- \\
 & \sigma^{(x)} & & 
\end{Bmatrix}
+ 
   \begin{Bmatrix}
 &  & \sigma^{(x)} & \sigma^{(x)}  \\
  & & & \\
\pi_+ & \sigma^{(x)} & \pi_-  &  \\
 & &  \pi_- & \pi_+ \\
 & \sigma^{(x)} & & 
\end{Bmatrix}, \hspace{0.3cm}
\label{toolongTLV1}
\end{eqnarray}
and 
\begin{equation}
   ( U^{(+1)}_{N_d,i,t})^\dag \sigma^{(x,-1)}_{i,t} U^{(+1)}_{N_d,i,t}= 
    \begin{Bmatrix}
 &  & \mathbb{1}   \\
  & & \\
\pi_+  & \pi_+  &  \\
 &  & \sigma^{(x)} 
\end{Bmatrix}
+
    \begin{Bmatrix}
 &  & \mathbb{1}   \\
  & & \\
\pi_-  & \pi_-  &  \\
 &  & \sigma^{(x)} 
\end{Bmatrix}
+   \begin{Bmatrix}
 &  & \sigma^{(x)}   \\
  & & \\
\pi_+  & \pi_-  &  \\
 &  & \sigma^{(x)}  
\end{Bmatrix}
+ 
  \begin{Bmatrix}
 &  & \sigma^{(x)}   \\
  & & \\
\pi_-  & \pi_+  &  \\
 &  & \sigma^{(x)} 
 
\end{Bmatrix}
.
\label{toolongTLV2}
\end{equation}
The action of the corresponding $(j=-1)$-automorphism on the $\sigma^{(x)}$ observable can be obtained with the aid of symmetry arguments. Note that in Eq.~\eqref{toolongTLV1} the columns from left to right correspond to $i-1$, $i$, $i+1$ and $i+2$, respectively, while the rows from bottom to top are labelled by $(t-1, j=+1)$, $(t, j=-1)$, $(t, j=+1)$, $(t+1, j=-1)$ and $(t+1, j=+1)$. Specifically, row $(t+1, j=-1)$ is kept empty in Eq.~\eqref{toolongTLV1} because the $(j=+1)$-automorphism does not act on it. In Eq.~\eqref{toolongTLV2} the columns from left to right correspond to $i-2$, $i-1$ and $i$, while the rows from bottom to top are labelled by $(t, j=-1)$, $(t, j=+1)$, $(t+1, j=-1)$ and $(t+1, j=+1)$. By taking the action of $U^{(-1)}_{N_d,i,t}$ on $\sigma^{(x,+1)}_{i,t}$ from Eq.~\eqref{toolongTLV2}, it is possible to show that the total automorphism acts on this observable to give a total of 64 terms. Similarly to the Q232 case, for any $j$ and $j'$ we have
\begin{equation}
     (U^{(j)}_{N_d,i,t})^\dag \sigma^{(z,j')}_{i,t} U^{(j)}_{N_d,i,t}= \sigma^{(z,j')}_{i,t} .
\end{equation}

 Lastly, the QTLV $j$-automorphism reads
\begin{equation}
   U_{t}^{\text{TLV}(j)}=  \exp \left\{\pm i\pi\sum_i\left[b^{(j)}_{i,t-1}{c^{(j)}_{i,t}}+ b^{(j)}_{i,t+1}{(c^{(j)}_{i-j,t}c^{(j)}_{i-2j,t}+c^{(j)}_{i-j,t}c^{(-j)}_{i,t}+c^{(-j)}_{i,t}c^{(j)}_{i-2j,t})}\right]\right\}
\end{equation}
with $b^{(j)}_{i,t}=[\mathbb{1}^{(j)}_{i,t}-\sigma^{(x,j)}_{i,t}]/2$ and $c^{(j)}_{i,t}=[\mathbb{1}^{(j)}_{i,t}-\sigma^{(z,j)}_{i,t}]/2$, while the total automorphism is given by
\begin{equation}
   U_{t}^{\text{TLV}}=  \exp \left\{\pm i\pi\sum_{i,j}\left[b^{(j)}_{i,t-1}{c^{(j)}_{i,t}}+ b^{(j)}_{i,t+1}{(c^{(j)}_{i-j,t}c^{(j)}_{i-2j,t}+c^{(j)}_{i-j,t}c^{(-j)}_{i,t}+c^{(-j)}_{i,t}c^{(j)}_{i-2j,t})}\right]\right\} .
\end{equation}

\section{Noise models for the simulation of quantum-error-correcting QCAs}

Since error-correcting performances are strongly dependent on noise models, and the latter can vary drastically from one platform to another, we consider 3 different noise models in this work. 

The first and simplest model is that of phenomenological incoherent bit-flip noise, which corresponds exactly to the noise model employed in the classical CA simulations. According to this model, before each QCA update the states of each time-$t$ cell (qubit) are acted upon by the incoherent channel 
\begin{equation}
    \varepsilon^p_{X,i,t} [\rho] = (1-p)\rho + p\sigma^{(x)}_{i,t} \rho \sigma^{(x)}_{i,t} , 
\end{equation}
where $\rho$ is the density matrix of the total system and $\sigma^{(x)}_{i,t} = \tilde{\sigma}^{(x)}_{i,t} \otimes_{l \neq i} \mathbb{1}_{l,t}$ is the Pauli-X operator from the observable algebra $\mathcal{A}_{i,t}$ localized in cell $i$ at time $t$ (notice that different cells are associated with different times). A full round of noise application is therefore given by $\varepsilon^p_{X,t}[\rho]\equiv \varepsilon^p_{X,n,t} \circ\ldots\circ\varepsilon^p_{X,1,t}  \circ\varepsilon^p_{X,0,t} [\rho]$. 

The second model corresponds to phenomenological coherent bit-flip noise, in which each QCA update is preceded by the action of the following channel on the states of each time-$t$ cell: 
\begin{equation}
   {\varepsilon}^\theta_{X,i,t} [\rho] =  e^{i \frac{\theta}{2}\sigma^{(x)}_{i,t}} \rho e^{-i \frac{\theta}{2}\sigma^{(x)}_{i,t}} .
   \label{coherent_ch}
\end{equation}
In Eq.~\eqref{coherent_ch}, $\sin^2(\theta/2)=p$, so that $\text{tr} \{ (|1_i\rangle \langle 1_i|\otimes_{l\neq i}|0_l\rangle \langle 0_l| ){\varepsilon}^\theta_{X,i} [\otimes_j|0_j\rangle \langle 0_j|] \} = p$. A full round of noise application is then described by
\begin{equation}
   \varepsilon^\theta_{X,t}[\rho]\equiv \varepsilon^\theta_{X,n,t} \circ\ldots\circ\varepsilon^\theta_{X,1,t}  \circ{\varepsilon}^\theta_{X,0,t} [\rho] =  e^{i \frac{\theta}{2}\sum_i\sigma^{(x)}_{i,t}} \rho e^{-i \frac{\theta}{2}\sum_i\sigma^{(x)}_{i,t}} .
\end{equation}

We refer to circuit noise when the noise is applied incoherently on all qubits that have been acted upon by a gate. If two gates act consecutively on the same qubit, the circuit noise will then lead to the ordering "circuit-noise-circuit-noise" on that qubit. Since our QCA quantum circuit is comprised of Toffoli and CNOT gates, we have two circuit-noise channels, one for each gate type. The Toffoli depolarizing noise channel reads
\begin{equation}
   {\varepsilon}^p_{T} [\rho] =  \left(1-\frac{4^3}{4^3-1}p\right)\rho + \frac{p}{4^3-1}\sum_{m,n,l \in \{0,x,y,z\}} 
 \sigma^{(l)}_{c1}\sigma^{(m)}_{c2}\sigma^{(n)}_{f} \rho \sigma^{(n)}_{f}\sigma^{(m)}_{c2}\sigma^{(l)}_{c1}
   \label{Toffoli_ch}
\end{equation}
and acts on the two control qubits, $c1$ and $c2$, as well as on the target qubit $f$ of the Toffoli gate. Here and in what follows, we denote $\sigma^{(0)}_{i,t} = \otimes_j\mathbb{1}_{j,t} \; \forall \, i$. Note that the coefficient in the first term on the right-hand side of Eq.~\eqref{Toffoli_ch} compensates for the inclusion of the identity in the sum in the second term. Similarly, the CNOT depolarizing noise channel is given by
\begin{equation}
   {\varepsilon}^p_{cN} [\rho] =  \left(1-\frac{4^2}{4^2-1}p\right)\rho + \frac{p}{4^2-1}\sum_{m,n \in \{0,x,y,z\}} 
 \sigma^{(m)}_{c}\sigma^{(n)}_{f} \rho \sigma^{(n)}_{f}\sigma^{(m)}_{c},
   \label{cnot_ch}
\end{equation}
acting on the control and target qubits of the CNOT gate. We denote a full round of circuit-noise application, covering all Toffoli and CNOT gates that constitute a QCA step, by the global channel $\tilde{\varepsilon}^p_{\Sigma}[\rho ]$.

\section{Quantum circuits for the simulation of Q232 and QTLV}

\begin{figure}
    \centering
    \includegraphics[width=\textwidth]{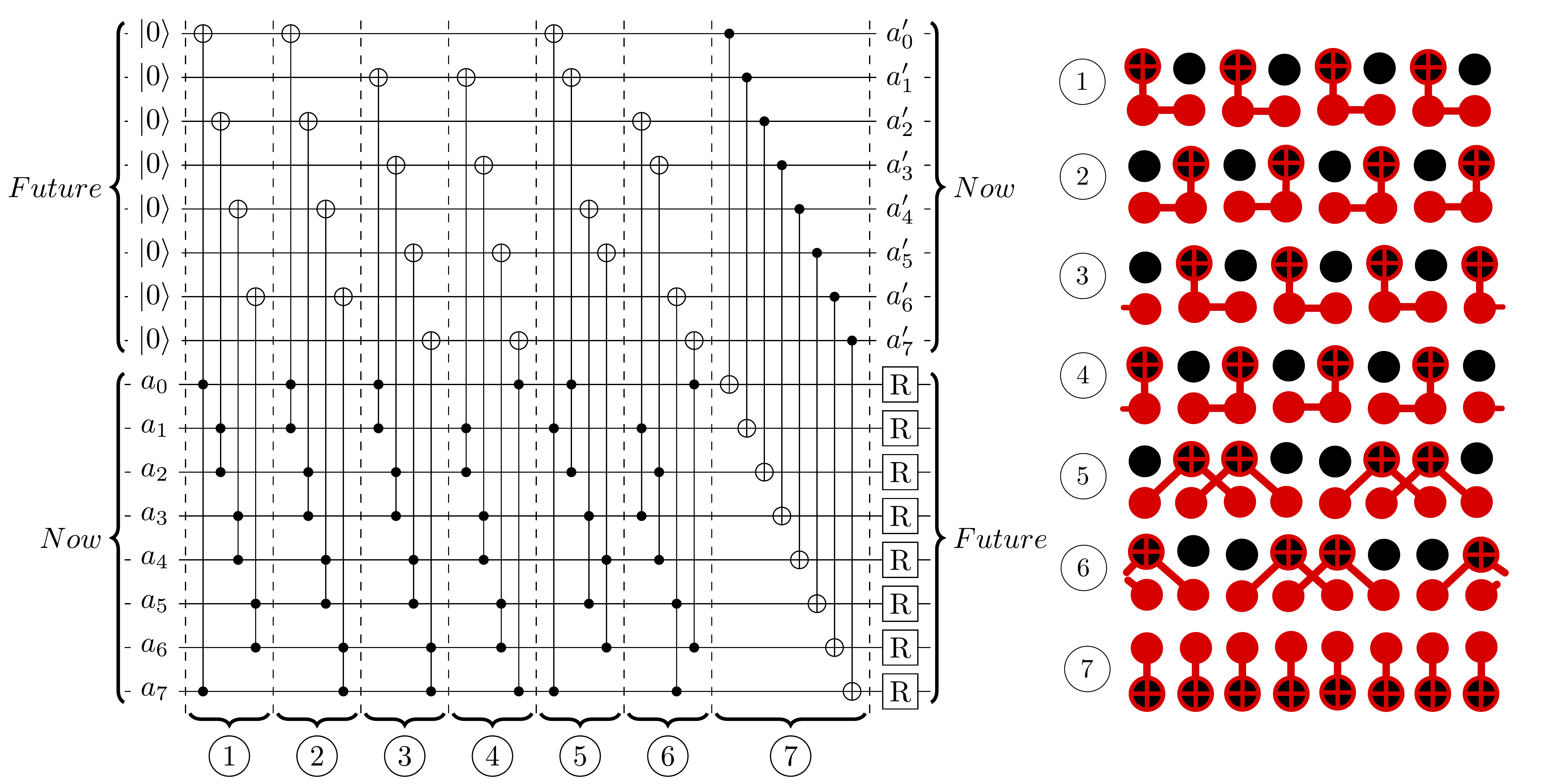}
 \caption{Q232 quantum circuit for application of one global rule. The Toffoli gates are ordered for parallel application, such that for all even lattice sizes $n$ one can realize one global update within 7 parallel gate applications both for Q232 and QTLV. The grouped gate applications are illustrated on a qubit array labeled by the encircled numbers. Gates can in principle be applied in parallel when none of their control or target qubits overlap. The geometrical arrangement of qubits (black circles) and gates (red) is diagrammatically displayed on the right.}
  \label{fig_circ1}
\end{figure}

The quantum circuit for the simulation of Q232, making use of fully parallelized Toffoli and CNOT gates, is shown in Fig.~\ref{fig_circ1}. Gate parallelization means that, at a given depth level (highlighted in Fig.~\ref{fig_circ1} by the numerals on the bottom of the circuit) each qubit, either control or target, is acted upon by at most one gate. To facilitate comprehension, the geometrical arrangements of those parallelized gates at each depth level are diagrammatically represented on the right side of Fig.~\ref{fig_circ1}, where qubits are represented as circles and gates are represented in red. We include 8 time-$t$ (or "\emph{Now}") data qubits labeled $a_0, a_1, a_2, a_3, a_4, a_5, a_6, a_7$, in which the logical state is encoded, and 8 additional time-$(t+1)$ (or "\emph{Future}") qubits initialized in the state $|0\rangle$. After a QCA step including all gates in Fig.~\ref{fig_circ1}, the "\emph{Future}" qubits, now storing the corrected logical state, will be relabelled to "\emph{Now}"; similarly, the "\emph{Now}" qubits $a_0, a_1, a_2, a_3, a_4, a_5, a_6, a_7$ will be reset to $|0\rangle$ and then relabelled to "\emph{Future}". The next QCA application will therefore use the top 8 qubits as controls and the bottom 8 qubits as targets (equivalently, a sequence of transversal SWAP gates could be used to interchange "\emph{Future}" and "\emph{Now}"). If each time slice contains an even number $n$ of qubits, grouping parallelized Toffoli gates so that each control and target qubit is acted upon by a gate exactly 3 times (to implement the majority) requires a depth of 6: if $n$ is a multiple of 4, the control qubits can be fully intercalated at depth levels 5 and 6, as shown in Fig.~\ref{fig_circ1}, otherwise 2 gates from a lower depth level have to be implemented at depth levels 5 and 6 (one gate per level) while some "next-nearest-neighbor-controlled" gates have to be implemented at that lower depth level. Since the CNOT gates are all transversal, their application accounts for a depth of 1.

\begin{figure}[h!]
    \centering
    \includegraphics[width=\textwidth]{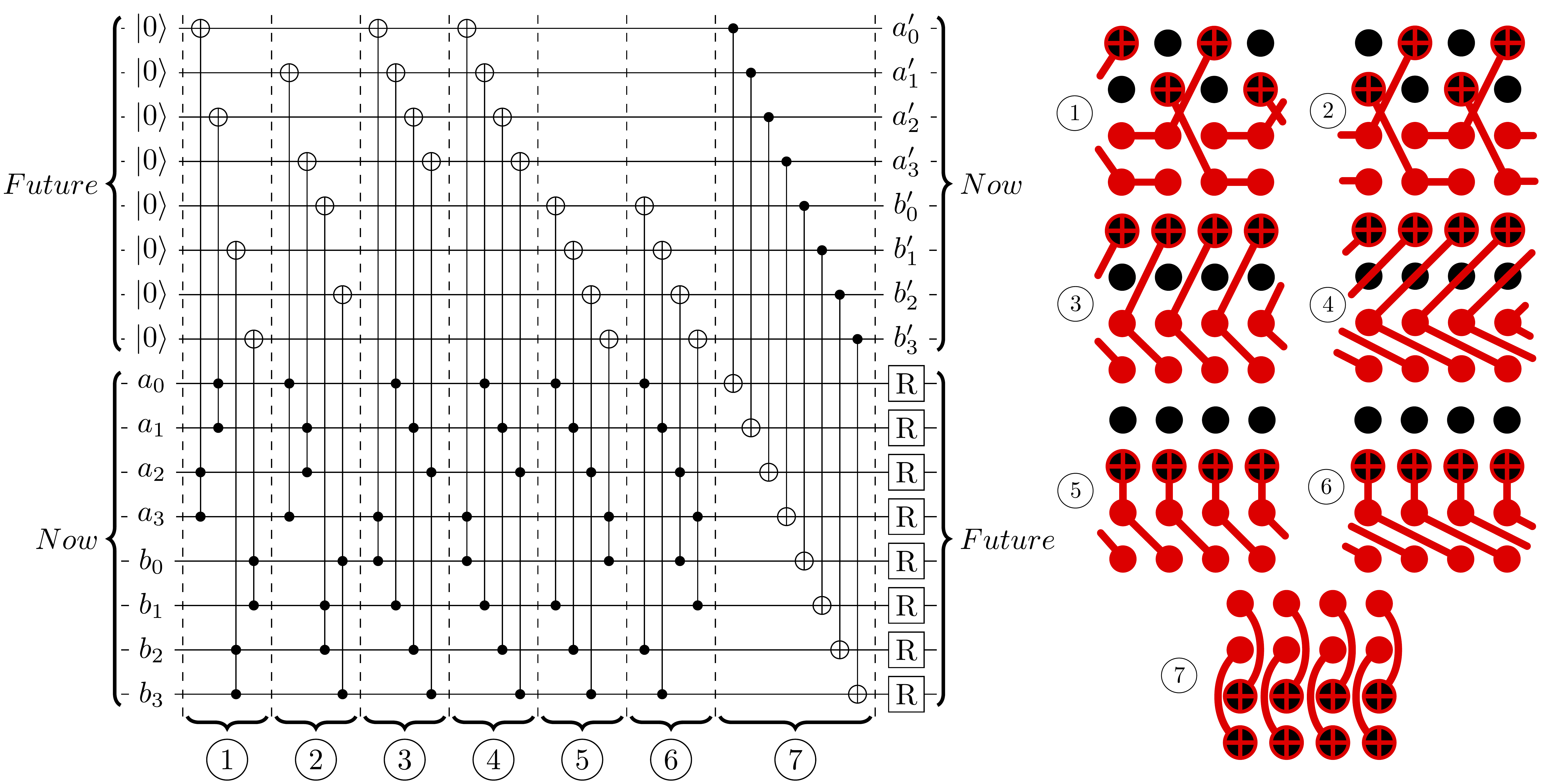}
 \caption{QTLV quantum circuit for application of one global rule. The geometrical arrangement of qubits (black circles) and gates (red) is diagrammatically displayed on the right.}
  \label{fig_circ2}
\end{figure}

The quantum circuit for the simulation of QTLV is shown in Fig.~\ref{fig_circ2},  where $j=1$ qubits are labelled as $a_0, a_1, a_2, a_3$ and $j=-1$ qubits as $b_0, b_1, b_2, b_3$. Explanations are similar to the ones provided above for the Q232. 

\begin{figure}[h!]
    \centering
    \includegraphics[width=0.6\textwidth]{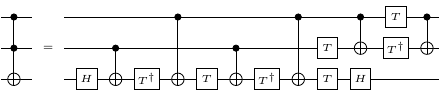}
    \caption{Toffoli gate decomposition in terms of single- and two-qubit gates.}
  \label{tof_id}
\end{figure}

It is worth noting that the potential unavailability of Toffoli gates in given experimental setups for implementations of those QCA circuits represents no limitation whatsoever, since a Toffoli gate can be decomposed in terms of single- and two-qubit gates in the well known form presented in Fig.~\ref{tof_id}~\citeS{cnot_decomposeS}.

\section{Majority gates vs Toffoli-gate decomposition}

\begin{figure}[h!]
    \centering
    \includegraphics[width=0.8\textwidth]{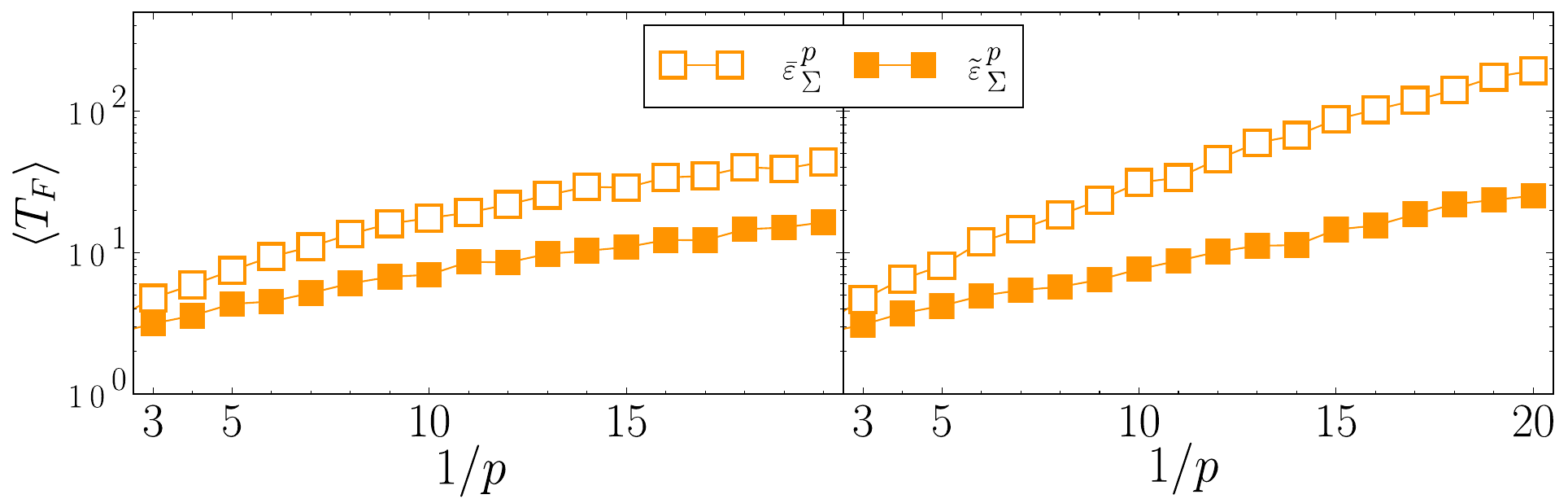}
    \caption{Performances of Q232 and QTLV under depolarizing gate noise when Toffoli gates (filled orange squares) or majority gates (void orange squares) are used. The Toffoli gates are afflicted by the noise channel \eqref{Toffoli_ch}, while the majority gates by \eqref{Majority_ch} [in both cases, the CNOT gates are afflicted by the channel \eqref{cnot_ch}], and the total noise channels are labelled $\tilde{\varepsilon}^p_\Sigma$ and $\bar{\varepsilon}^p_\Sigma$, respectively. It is possible to see that employing native majority gates, if available, instead of Toffoli ones can drastically increase the performance of QTLV under this noise model.}
  \label{Tof_vs_maj}
\end{figure}

In the main text we introduced a decomposition of the quantum-local-majority-voting unitary in terms of Toffoli gates and a disentangling CNOT gate. Depending on the lattice arrangement, this unitary generates either the Q232 or QTLV global updates. It is worth noting, however, that the Toffoli unitary, due to its non-Clifford nature, maps Pauli operators into a tensor product of Pauli operators and controlled Pauli operators. This mapping leads to error patterns that violate self-duality when considering gate-based noise models, as in Eqs.~\eqref{Toffoli_ch} and \eqref{cnot_ch}: the error configurations afflicting the logical zero and one states in superposition are different at the end of a QCA update. Furthermore, the minimal-depth weaving of Toffoli gates presented in Figs.~\ref{fig_circ1} and \ref{fig_circ2} leads to propagated errors correlated both in time (i.e., between different-time registers) and (non-locally) in space. The decomposition of the quantum-local-majority-voting unitary into noisy Toffoli gates therefore considerably affects the working mechanism of our QCAs, since error propagation violates both locality preservation and self-duality. One way of circumventing this issue is by assuming the existence of a native majority gate with the net effect of 3 Toffoli gates (the CNOT should then be additionally applied for decoupling). Without the need for sub-partition into smaller gates or gate weaving, the noisy majority gate allows for conservation of the basic symmetries of the QCAs and therefore can improve the performance of the considered systems, as shown in Fig.~\ref{Tof_vs_maj}, where we considered the noise channel
\begin{equation}
   {\varepsilon}^p_{M} [\rho] =  \left(1-\frac{4^4}{4^4-1}p\right)\rho + \frac{p}{4^4-1}\sum_{m,n,k,l \in \{0,x,y,z\}} 
 \sigma^{(k)}_{c1}\sigma^{(l)}_{c2}\sigma^{(m)}_{c3}\sigma^{(n)}_{f} \rho \sigma^{(n)}_{f}\sigma^{(m)}_{c3}\sigma^{(l)}_{c2} \sigma^{(k)}_{c1}.
   \label{Majority_ch}
\end{equation}
for the action of the majority gate on the three control qubits, $c1$, $c2$ and $c3$, as well as on the target qubit $f$.

\section{Full quantum error correction through QCA concatenation}

It can be argued that, since Q232 and QTLV protect the same code space as the repetition code, the presented QCA designs are actually not performing full quantum error correction. The quantum repetition code leaves the quantum information fully exposed to phase flips, since single physical-qubit phase flips (or, more generally, any odd number of such flips) are also logical operations. Concatenation of repetition codes in different bases, however, enables for the protection of quantum information against both bit and phase flips: the $[[mn, 1, \text{min}(m,n)]]$ Shor code is a quantum code whose logical states are given by $|0\rangle_L =|+\rangle_B^{\otimes m}$ and $|1\rangle_L =|-\rangle_B^{\otimes m}$, in which $|+\rangle_B =\frac{1}{\sqrt{2}}(|0\rangle^{\otimes n}+|1\rangle^{\otimes n})$ and $|-\rangle_B =\frac{1}{\sqrt{2}}(|0\rangle^{\otimes n}-|1\rangle^{\otimes n})$ are $n$-qubit repetition-code logical states of each of the $m$ blocks of the concatenation. Under the effect of global data collection, processing and correction, as in the case of minimum-weight perfect matching, logical bit flips in the $[[mn, 1, \text{min}(m,n)]]$ Shor code are induced by block-state bit flips afflicting the majority of the $m$ blocks (i.e., odd numbers of physical phase flips within each block), while logical phase flips are induced by phase-flipping an odd number of blocks, each of which is caused by physical bit flips in the majority of the $n$ qubits in each block. The asymmetry in the error-induction mechanisms for logical phase and bit flips in the Shor code leads to different rates/probabilities of appearance of each of these errors in a given logical state. Nonetheless, both types of errors can be corrected by applying the repetition-code routine both within each block and amongst blocks. 

\begin{figure}[h!]
    \centering
    \includegraphics[width=0.5\textwidth]{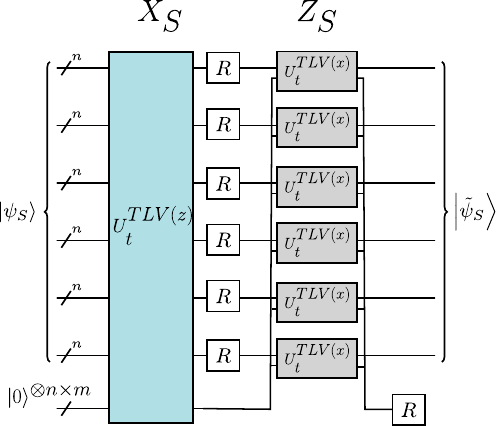}
    \caption{Schematic representation of the concatenated quantum TLV (CQTLV). The (present-time) logical state of the $[[mn, 1, \text{min}(m,n)]]$ Shor code space, $|\psi_S\rangle$ is encoded in $m$ $n$-qubit blocks (horizontal black lines), each of which is in a state $|\pm\rangle_B =\frac{1}{\sqrt{2}}(|0\rangle^{\otimes n}\pm |1\rangle^{\otimes n})$ when $|\psi_S\rangle = |0\rangle_L$ or $|\psi_S\rangle = |1\rangle_L$, respectively. A future-time register with $mn$ qubits also organized block-wise is provided, so that the higher-level phase-flip-correcting QTLV, represented as a cyan box, can move the block-locally corrected logical state upwards in time and sequentially decouple the present and future registers. After decoupling, the present-time register contains irrelevant information, which can then be reset to physical $|0\rangle$ states (white boxes). The future register, now containing a phase-flip-corrected logical state, then undergoes a unitary entangling operation block-wise, namely the bit-flip-correcting QTLV (grey boxes), which moves the information in the blocks back to the present register while locally correcting bit-flip errors. These block operations also contain a decoupling step between the two registers, so that the future register, which now carries junk information, can be reset before a new QCA update starts. The result of one such QCA update is a logical state $|\tilde{\psi}_S\rangle$ which differs from $|\psi_S\rangle$  up to local bit-flip corrections and block-local phase flip corrections.}
  \label{CQTLV_scheme}
\end{figure}

A similar approach can be adopted for our QCA designs: concatenating bit-flip-correcting QCAs with phase-flip-correcting ones generates a full QEC protocol. One can run autonomous, local (i.e., locality-preserving) and measurement-free bit-flip correction on each of the $m$ blocks of the quantum state with QTLV, for example, while performing a higher-order phase-flip-correcting QTLV whose majority is taken amongst block states. For the current discussion, we abstract ourselves from specific universal-gate-set-based quantum-circuit recastings of QCAs, and instead assume the availability of a proper QCA architecture with a majority gate that can be applied simultaneously on all 3 control qubits and the target qubit. We postpone the description of the gate decomposition and quantum circuit implementation of such concatenation to a posterior work. The proposed scheme is shown in Fig.~\ref{CQTLV_scheme}, where the present-time logical state $|\psi_S\rangle$ is encoded into $m$ blocks (solid horizontal lines) of $n$ qubits each, and is fed into the concatenated QTLV alongside a future-time logical state $|0\rangle^{\otimes mn}$. The higher-order phase-flip-correcting sub-protocol is implemented by applying QTLV on the blocks with majority gates acting on 
3 time-$t$ and one time-$(t+1)$ block state according to the truth table:
\begin{table}[h!]
	\centering
	\begin{tabular}{l c c c|c c c r}
	\multicolumn{4}{c}{Input} & \multicolumn{4}{c}{Output}\\
	\hline
	$|+\rangle_B^t$ & $|+\rangle_B^t$ & $|+\rangle_B^t$ & $|0\rangle_B^{t+1}$ & $|+\rangle_B^t$ & $|+\rangle_B^t$ & $|+\rangle_B^t$ & $|+\rangle_B^{t+1}$ \\
	$|+\rangle_B^t$ & $|+\rangle_B^t$ & $|-\rangle_B^t$ & $|0\rangle_B^{t+1}$ & $|+\rangle_B^t$ & $|+\rangle_B^t$ & $|-\rangle_B^t$ & $|+\rangle_B^{t+1}$ \\
	$|+\rangle_B^t$ & $|-\rangle_B^t$ & $|+\rangle_B^t$ & $|0\rangle_B^{t+1}$ & $|+\rangle_B^t$ & $|-\rangle_B^t$ & $|+\rangle_B^t$ & $|+\rangle_B^{t+1}$ \\
	$|+\rangle_B^t$ & $|-\rangle_B^t$ & $|-\rangle_B^t$ & $|0\rangle_B^{t+1}$ & $|+\rangle_B^t$ & $|-\rangle_B^t$ & $|-\rangle_B^t$ & $|-\rangle_B^{t+1}$ \\
	$|-\rangle_B^t$ & $|+\rangle_B^t$ & $|+\rangle_B^t$ & $|0\rangle_B^{t+1}$ & $|-\rangle_B^t$ & $|+\rangle_B^t$ & $|+\rangle_B^t$ & $|+\rangle_B^{t+1}$ \\
	$|-\rangle_B^t$ & $|+\rangle_B^t$ & $|-\rangle_B^t$ & $|0\rangle_B^{t+1}$ & $|-\rangle_B^t$ & $|+\rangle_B^t$ & $|-\rangle_B^t$ & $|-\rangle_B^{t+1}$ \\
	$|-\rangle_B^t$ & $|-\rangle_B^t$ & $|+\rangle_B^t$ & $|0\rangle_B^{t+1}$ & $|-\rangle_B^t$ & $|-\rangle_B^t$ & $|+\rangle_B^t$ & $|-\rangle_B^{t+1}$ \\
	$|-\rangle_B^t$ & $|-\rangle_B^t$ & $|-\rangle_B^t$ & $|0\rangle_B^{t+1}$ & $|-\rangle_B^t$ & $|-\rangle_B^t$ & $|-\rangle_B^t$ & $|-\rangle_B^{t+1}$ \\
	\hline
	$|+\rangle_B^t$ & $|+\rangle_B^t$ & $|+\rangle_B^t$ & $|1\rangle_B^{t+1}$ & $|+\rangle_B^t$ & $|+\rangle_B^t$ & $|+\rangle_B^t$ & $|-\rangle_B^{t+1}$ \\
	$|+\rangle_B^t$ & $|+\rangle_B^t$ & $|-\rangle_B^t$ & $|1\rangle_B^{t+1}$ & $|+\rangle_B^t$ & $|+\rangle_B^t$ & $|-\rangle_B^t$ & $|-\rangle_B^{t+1}$ \\
	$|+\rangle_B^t$ & $|-\rangle_B^t$ & $|+\rangle_B^t$ & $|1\rangle_B^{t+1}$ & $|+\rangle_B^t$ & $|-\rangle_B^t$ & $|+\rangle_B^t$ & $|-\rangle_B^{t+1}$ \\
	$|+\rangle_B^t$ & $|-\rangle_B^t$ & $|-\rangle_B^t$ & $|1\rangle_B^{t+1}$ & $|+\rangle_B^t$ & $|-\rangle_B^t$ & $|-\rangle_B^t$ & $|+\rangle_B^{t+1}$ \\
	$|-\rangle_B^t$ & $|+\rangle_B^t$ & $|+\rangle_B^t$ & $|1\rangle_B^{t+1}$ & $|-\rangle_B^t$ & $|+\rangle_B^t$ & $|+\rangle_B^t$ & $|-\rangle_B^{t+1}$ \\
	$|-\rangle_B^t$ & $|+\rangle_B^t$ & $|-\rangle_B^t$ & $|1\rangle_B^{t+1}$ & $|-\rangle_B^t$ & $|+\rangle_B^t$ & $|-\rangle_B^t$ & $|+\rangle_B^{t+1}$ \\
	$|-\rangle_B^t$ & $|-\rangle_B^t$ & $|+\rangle_B^t$ & $|1\rangle_B^{t+1}$ & $|-\rangle_B^t$ & $|-\rangle_B^t$ & $|+\rangle_B^t$ & $|+\rangle_B^{t+1}$ \\
	$|-\rangle_B^t$ & $|-\rangle_B^t$ & $|-\rangle_B^t$ & $|1\rangle_B^{t+1}$ & $|-\rangle_B^t$ & $|-\rangle_B^t$ & $|-\rangle_B^t$ & $|+\rangle_B^{t+1}$ 
	\end{tabular}
\end{table}

QTLV will therefore be composed of $4n$-qubit higher-level majority gates, which can only be seen as locality preserving with respect to block states. The global phase-flip-correcting update is represented in Fig.~\ref{CQTLV_scheme} as $U^{TLV(z)}_t$. The standard QTLV, as introduced in the main text, acts within each block to correct bit flips in the states $|\pm\rangle^t_B$, and their global unitary is represented in Fig.~\ref{CQTLV_scheme} by $U^{TLV(x)}_t$. It is important to notice that each of the sub-protocols involves at least two registers, corresponding to two time slices, yet implementing the sub-protocols in sequence with reset operations in between, which allow for the use of the same set of registers for both sub-protocols. 

\begin{figure}[h!!]
    \centering
    \includegraphics[width=0.6\textwidth]{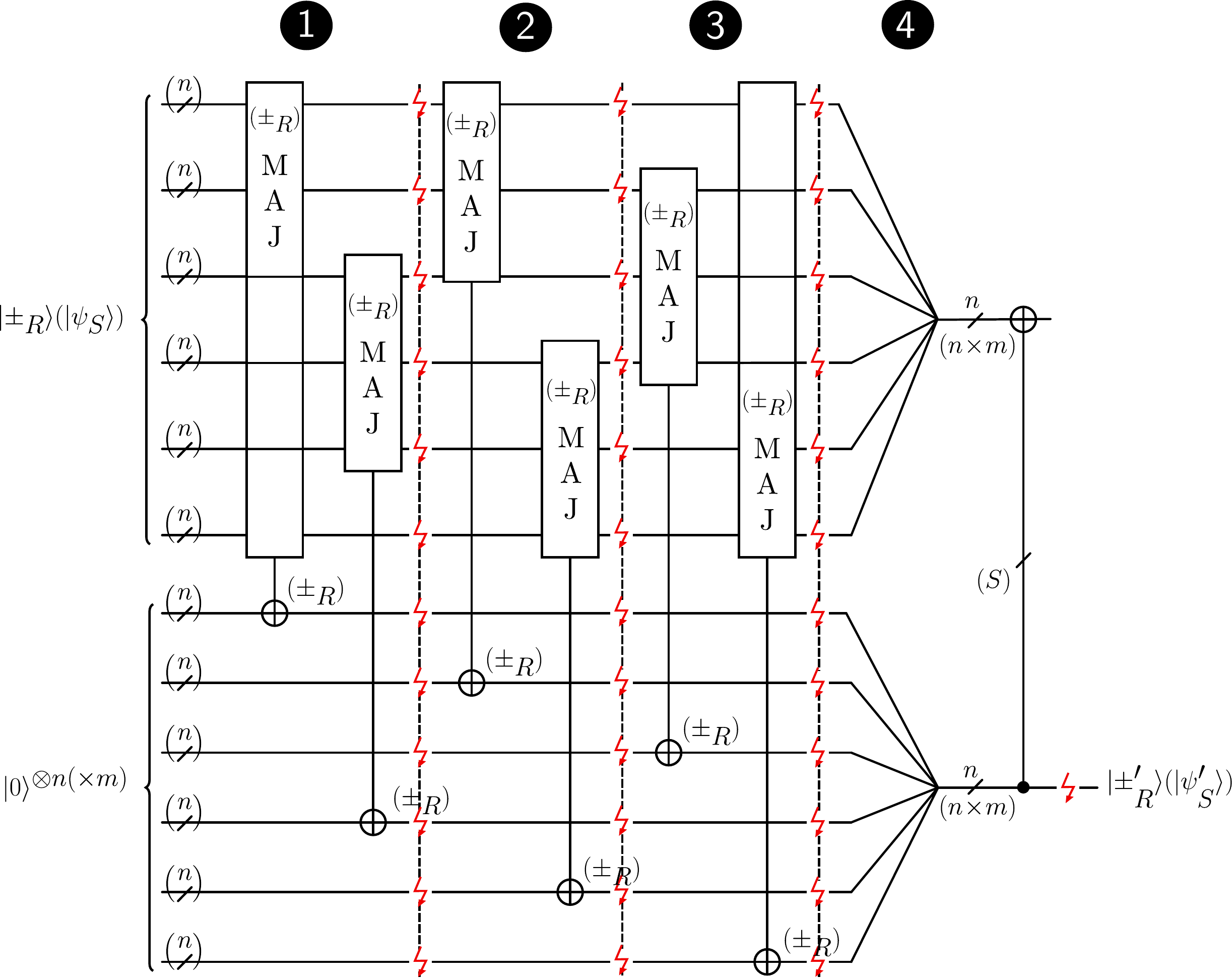}
    \caption{Schematic representation of noise application in the CQTLV. The scheme is identical for both the $U^{TLV(x)}_t$ and $U^{TLV(z)}_t$ unitaries, with quantities associated to the latter represented in parentheses. In the present-time register, an $n$-qubit block state $|\pm_R\rangle$ (or a $mn$ qubit logical state $|\psi_S\rangle$) is fed into the QTLV with an additional future-register of $|0\rangle$ states. The majority unitaries composing $U^{TLV(x)}_t$ act on physical qubits in the $\{|0\rangle , |1\rangle \}$ basis, while for $U^{TLV(z)}_t$ they act on the repetition-code states in the blocks, whose basis is $\{|+\rangle_B , |-\rangle _B\}$. Each depth-2 layer of majority gates acting on all present-time qubits is followed by an application of physical bit and phase flips on all qubits of the system with probability $p$. After all majority gates are implemented, a round of decoupling is performed, followed by another application of noise. For the phase-flip-correcting unitary, the decoupling is performed by the Shor-code logical CNOT gate (S), which can be implemented transversally as CNOT gates with swapped controls and targets. Note that the logical Shor-code CNOT is transversal but the roles of controls and targets are interchanged (as we basically apply transversal CNOT between blocks which are in the repetition code X basis, thus the CNOTs are reversed). For the bit-flip correcting unitary this is simply a transversal CNOT.}
  \label{CQTLV_error}
\end{figure}

\begin{figure}[h!!]
    \centering
    \includegraphics[width=0.5\textwidth]{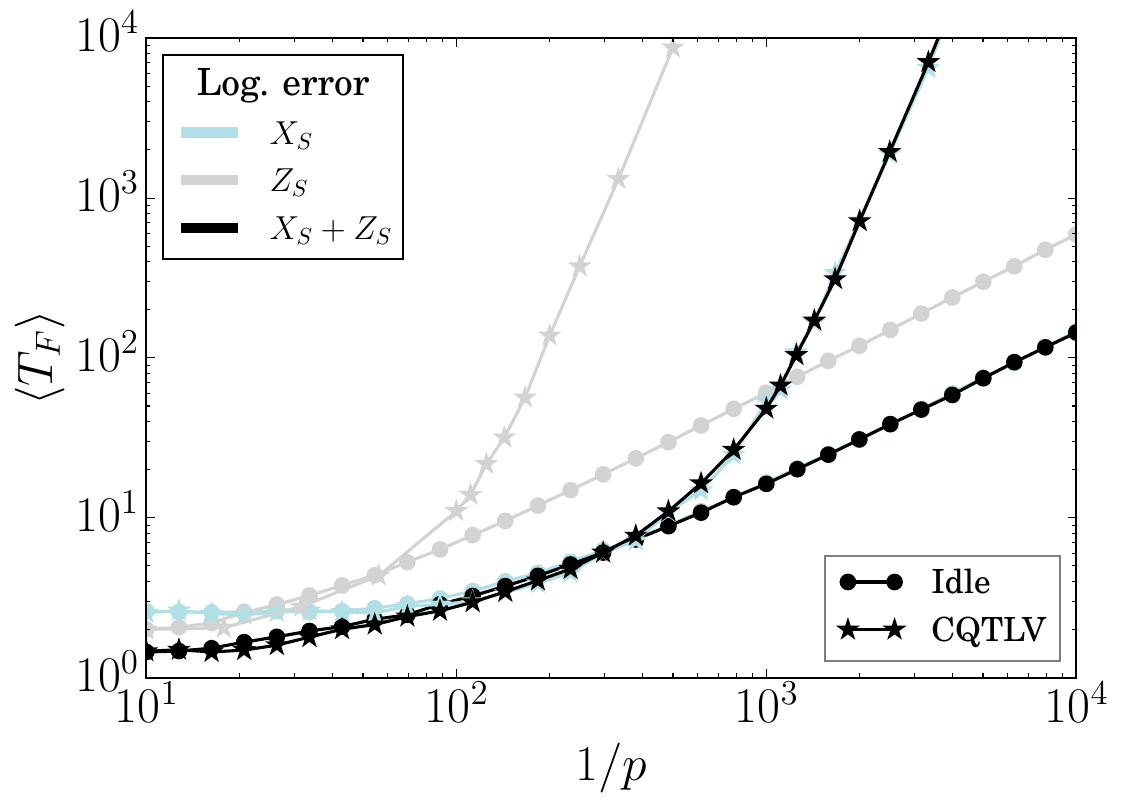}
    \caption{Flip times for the CQTLV-protected and unprotected $[[mn, 1, \text{min}(m,n)]]$ Shor logical state with $m=n=12$. Logical bit flips and phase flips are counted separately, leading to 2 flip-time profiles for each investigated scenario: the cyan and the grey curves register logical bit and phase flips, respectively. The actual flip times are taken as the minimum amongst the bit-flip- and phase-flip-based flip times, which for both protected and unprotected logical states are defined by logical bit flips at sufficiently low physical error probabilities. Data points corresponding to the unprotected logical state and to the state protected by CQTLV are represented by circles and stars, respectively. An improvement in the actual flip time of the logical state can be achieved for physical error probabilities smaller than $p\approx 2.5\times 10^{-3}$ when the noise model shown in Fig.~\ref{CQTLV_error} is applied.}
  \label{CQTLV_benchmark}
\end{figure}

\begin{figure}[h!!]
    \centering
    \includegraphics[width=0.6\textwidth]{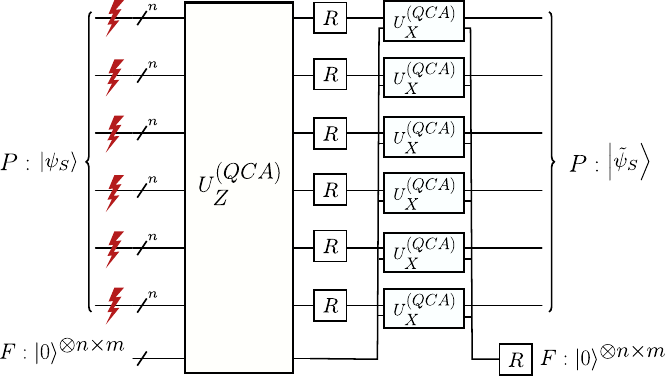}
    \caption{Schematic representation of the simplified noise model for CQTLV.  Single-qubit bit- and phase-flip noise is applied on each physical qubit of the system incoherently with probablity $p$ before each QCA cycle, as marked by the lightning signs. All the gates are then implemented in a noise-free fashion.}
  \label{CQTLV_error2}
\end{figure}

\begin{figure}[h!!]
    \centering
    \includegraphics[width=0.5\textwidth]{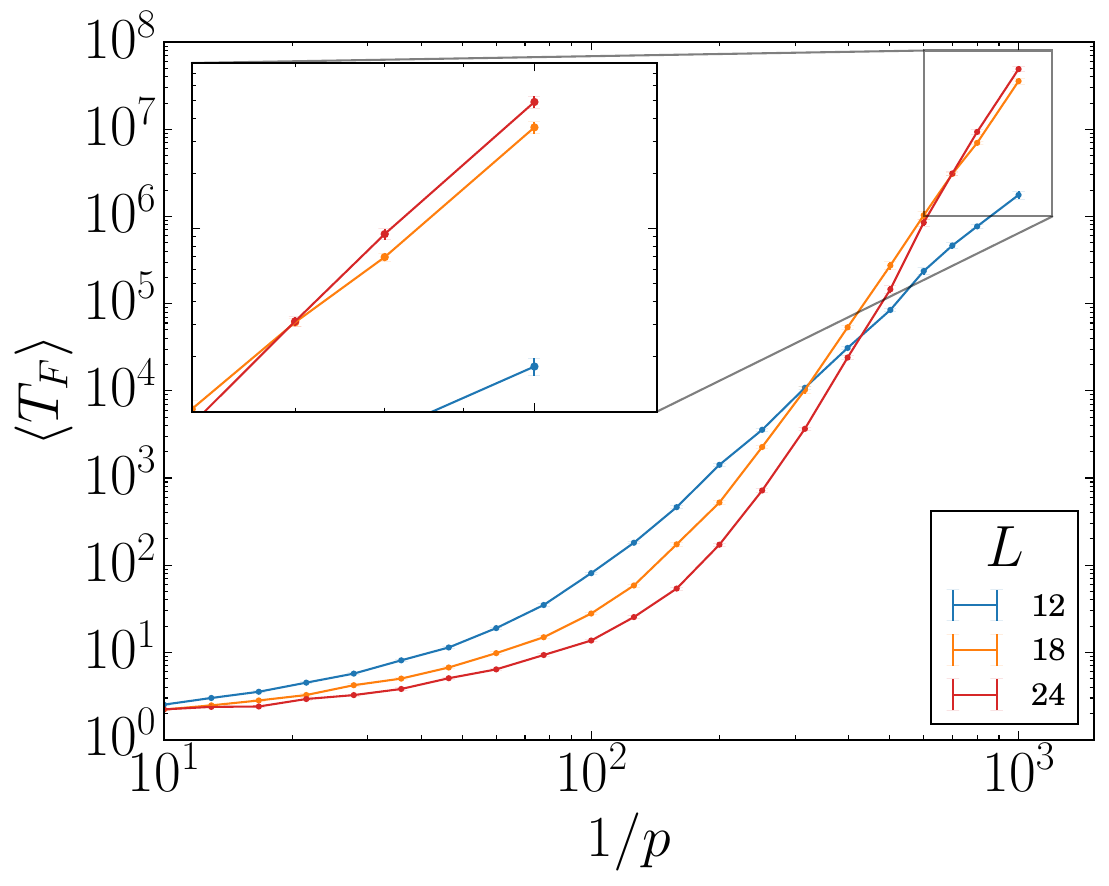}
    \caption{Flip times for the CQTLV-protected $[[mn, 1, \text{min}(m,n)]]$ Shor logical state afflicted by the noise model shown in Fig.~\ref{CQTLV_error2}. The actual flip times are taken as the minimum amongst the bit-flip- and phase-flip-based flip times.  Data points corresponding to different investigated system sizes (with $m=n=L$) are represented by color-coded circles. In the range $1/p \gsim 10^3$ it is possible to see that the flip times increase with the system size.}
  \label{CQTLV_benchmark2}
\end{figure}

We study the performance of our concatenated QTLV, CQTLV, in terms of flip times for different probabilities of physical bit- and phase-flip errors, both occurring at the same rate. Our noise model for this analysis implements bits and phase flips on each physical qubit with a probability $p$ after each gate operation involving all time-$t$ qubits. The noise application schedule is represented schematically in Fig.~\ref{CQTLV_error}, where the dashed lines with lightning signs mark the application of noise. As a comparison, we study the quantum-information degradation of the unprotected logical state $|0\rangle_L$ under application of the same type of noise with same probability $p$ at the same levels of depth, so that an equivalent flip time could be defined for the unprotected, i.e.~not actively corrected, encoded logical state. The goal of such comparison is to enable judgement on whether the CQTLV could provide any advantage in protecting quantum information against both types of Pauli noise and, if so, under which conditions this is possible or even the most effective. Our simulations were performed classically through error book-keeping and Pauli-noise propagation, and logical errors were counted as soon as the aforementioned global error-inducing conditions were met. Since the error-induction mechanisms for logical bit and phase flips differ considerably, flip times were estimated separately for each logical-error condition, as well as for both combined. Our results are shown in Fig.~\ref{CQTLV_benchmark}. It is possible to see that for $p\lesssim 10^{-3}$ the flip times registered for CQTLV for both logical-flip conditions surpass the corresponding quantities for the unprotected code (i.e., the logical-information lifetime). Such logical-information lifetime improvement under the action of CQTLV serves as a proof of principle that QCAs can be used for full QEC, as stated in the main text. It is worth noting, however, that more intricate designs and exhausting benchmarking are necessary to bring QCA-based full QEC to a level at which competitiveness against global QEC codes under experimentally realistic conditions can be considered. As a last remark, we would like to clarify that even for relatively small $p$ values there seems to be no flip-time improvement as either $m$ or $n$ increase (in fact, the opposite trend was observed for a few investigated sizes), what might hint at an important locality-related constraint on the concatenation-based QCA design (e.g., the number of faulty configurations increases too rapidly compared to the number of correctible ones, since the latter is limited by the performance of quasi-local corrections on subsets of qubits of a certain maximal size). Nonetheless, investigation of a simpler error model in which noise is applied only once before each QCA cycle (see Fig.~\ref{CQTLV_error2}) reveals that for sufficiently small $p$ a performance improvement with increasing system size is possible, as shown in Fig.~\ref{CQTLV_benchmark2}. 

\clearpage

\bibliographystyleS{prsty}
\bibliographyS{litQCAS}

\end{document}